%% file: main.tex
\documentclass{article}

\usepackage{arxiv}

\usepackage[utf8]{inputenc} % allow utf-8 input
\usepackage[T1]{fontenc}    % use 8-bit T1 fonts
\usepackage{hyperref}       % hyperlinks
\usepackage{url}            % simple URL typesetting
\usepackage{booktabs}       % professional-quality tables
\usepackage{amsfonts}       % blackboard math symbols
\usepackage{nicefrac}       % compact symbols for 1/2, etc.
\usepackage{microtype}      % microtypography
\usepackage{lipsum}		% Can be removed after putting your text content
\usepackage{graphicx}
\usepackage{doi}
\usepackage{subcaption}
\usepackage{proba}
\usepackage{bm}
\usepackage{enumitem}
\usepackage{xcolor}
\usepackage{afterpage}
\usepackage[authoryear]{natbib}
\usepackage[symbol]{footmisc}
\usepackage{bigints}
\usepackage{multirow}

\usepackage{setspace}
\usepackage[capitalise, noabbrev]{cleveref}

\usepackage{subfiles} % Best loaded last in the preamble

\title{A Posteriori Risk Classification and Ratemaking with Random Effects in the Mixture-of-Experts Model}

\author{ 
    Spark C.~Tseung\thanks{Department of Statistical Sciences, University of Toronto. Ontario Power Building, 700 University Avenue, 9th Floor, Toronto, ON M5G 1Z5, Canada. Email addresses: \texttt{spark.tseung@mail.utoronto.ca} (Spark C.~Tseung), \texttt{ianweng.chan@mail.utoronto.ca} (Ian Weng Chan), \texttt{andrei.badescu@utoronto.ca} (Andrei L.~Badescu), \texttt{sheldon.lin@utoronto.ca} (X.~Sheldon Lin).}\\ 
	\And
	{Ian Weng Chan$^*$} \\
	\And
	{Tsz Chai Fung\thanks{Department of Risk Management and Insurance, Georgia State University. 35 Broad Street NW, Atlanta, GA 30303, United States. Email address: \texttt{tfung@gsu.edu}.}} \\
	\And
	{Andrei L.~Badescu$^*$} \\
    \And
	{X.~Sheldon Lin$^*$} \\
}

\date{September 19, 2022}

\renewcommand{\headeright}{}
\renewcommand{\shorttitle}{A Posteriori Risk Classification and Ratemaking with Random Effects in the Mixture-of-Experts Model}

\hypersetup{
pdftitle={A Posteriori Risk Classification and Ratemaking with Random Effects in the Mixture-of-Experts Model},
pdfsubject={},
pdfauthor={Spark C.~Tseung, Ian Weng Chan, Tsz Chai Fung, Andrei L.~Badescu and Sheldon X.~Lin},
pdfkeywords={Risk Classification, Ratemaking, Mixture of Experts, Random Effects, Variational Inference},
}

\begin{document}
\maketitle

\begin{abstract}
	A well-designed framework for risk classification and ratemaking in automobile insurance is key to insurers' profitability and risk management, while also ensuring that policyholders are charged a fair premium according to their risk profile. In this paper, we propose to adapt a flexible regression model, called the \emph{Mixed LRMoE}, to the problem of \emph{a posteriori} risk classification and ratemaking, where policyholder-level random effects are incorporated to better infer their risk profile reflected by the claim history. We also develop a stochastic variational Expectation-Conditional-Maximization algorithm for estimating model parameters and inferring the posterior distribution of random effects, which is numerically efficient and scalable to large insurance portfolios. We then apply the Mixed LRMoE model to a real, multiyear automobile insurance dataset, where the proposed framework is shown to offer better fit to data and produce posterior premium which accurately reflects policyholders' claim history.
\end{abstract}

% keywords can be removed
\keywords{Risk Classification \and Ratemaking  \and Mixture of Experts \and Random Effects \and Variational Inference}

\section{Introduction} \label{sec:Introduction}

A well-designed framework for risk classification and ratemaking in automobile insurance is key to insurers' profitability and risk management, while also ensuring that policyholders are charged a fair premium according to their risk profile. For a new policyholder, risk classification and ratemaking are usually done on an \emph{a priori} basis, whereby the insurer only knows a set of the policyholder's covariates such as age, gender, vehicle specifications, etc. As time goes by, the insurer gains additional, up-to-date insights into the policyholder's risk profile from their claim history,  including frequency and severity, which leads to \emph{a posteriori} risk classification and ratemaking.

The use of claim history for \emph{a posteriori} risk classification and ratemaking is a classical problem which has been studied in depth in the actuarial literature. Early works in credibility theory, such as \cite{buhlmann1967experience}, \cite{norberg1979credibility} and \cite{buhlmann2005course}, assume some common parameters underlying the distribution of insurance losses. One uses the observed claim history to infer the posterior distribution of the parameters, which then yields the posterior distribution of future losses given the history. From a practical perspective, the Bonus-Malus System (BMS) is perhaps one of the most widely used approaches, see e.g.~\cite{lemaire1995bonus} and \cite{denuit2007actuarial}. Based on the claim history (typically the number of claims in the year prior to policy renewal), policyholders are (re-)classified into one of a number of pre-specified risk classes according to certain transition rules, whereby each risk class corresponds to a premium relativity which reflects the level of risk. However, in their classical formulation, neither credibility theory nor BMS considers covariate information, which is usually deemed as important indicators of policyholders' risk characteristics. To this end, there has been an abundance of literature that aims to apply more sophisticated statistical models, which typically involve a regression component, to the problem of \emph{a posteriori} risk classification and ratemaking. Most notably, random effects have been a popular choice for modelling the temporal dependence between past and future claim behaviour. For example, many authors have considered adding random effects in Generalized Linear Models (GLM), which results in Generalized Linear Mixed Models (GLMM), see e.g.~\cite{dionne1989generalization}, \cite{dionne1992automobile}, \cite{pinquet1998designing}, \cite{frangos2001design} and \cite{boucher2006fixed}, whereby the posterior distribution of random effects given claim history is used for prediction. Another important consideration is the dependence structure between multiple coverages which is common in automobile insurance, see e.g.~\cite{pinquet1998designing}, \cite{gomez2008univariate}, \cite{boucher2009number}, \cite{gomez2016bivariate} and \cite{tzougas2021multivariate} for using shared random effects to model such dependence. Besides, while some works mainly focus on claim frequency alone, many researchers have also attempted to incorporate claim severity and its dependence structure with frequency, for example, \cite{ni2014bonus}, \cite{park2018does}, \cite{oh2020bonus} and \cite{oh2021designing}. Furthermore, to overcome certain restrictive assumptions in GLM, finite mixture models have recently become popular in \emph{a posteriori} risk classification and ratemaking for more flexible and accurate modelling of claim frequency and severity, as used in \cite{bermudez2012finite}, \cite{tzougas2014optimal},  \cite{tzougas2018bonus} and \cite{tzougas2021multivariate}.

In this paper, we propose to apply a flexible regression model, called the \emph{Mixed LRMoE}, to the problem of \emph{a posteriori} risk classification and ratemaking. Compared with existing approaches to this problem, our proposed method enjoys several distinct advantages, such as an intuitive and interpretable model structure (see \cref{sec:Mixed-LRMoE-overview}), the flexibility to model any mixed effects model (see \cref{sec:Mixed-LRMoE-dense}), and superior performance in goodness-of-fit and adequacy in \emph{a posteriori} risk classification and ratemaking compared with benchmark models (see \cref{sec:RealDataAnalysis}). The Mixed LRMoE as a general modelling framework has recently been introduced in \cite{Fung2022MixedLRMoETheory} as an extension to the Logit-weighted Reduced Mixture-of-Experts (LRMoE) model. The latter was first developed in \cite{fung2019class} and has subsequently been applied to various insurance modelling problems such as correlated claim frequencies and reporting delay (see \cref{sec:LRMoE-overview} for an overview). In order to adapt to the problem of \emph{a posteriori} risk classification and ratemaking, we propose to add policyholder-level random effects in a multiyear portfolio which results in the Mixed LRMoE. Similar to many papers cited above, the addition of random effects introduces dependence between observations across multiple policy years of the same policyholder, from which the posterior distribution of random effects is inferred and then utilized for \emph{a posteriori} risk classification and ratemaking. Our work also intersects with mixture model-based approaches such as \cite{tzougas2021multivariate}, in that the Mixed LRMoE model allows for more flexible and accurate modelling of the loss distribution compared with classical regression models such as GLM. In the broader class of general mixture-of-experts (MoE) models, our work is closely related to \cite{yau2003finite}, \cite{ng2007extension} and \cite{ng2014mixture}, where random effects are also incorporated to account for heterogeneity observed in real data. However, the Mixed LRMoE presented in this paper has an arguably simpler model structure. A detailed discussion on various properties of the Mixed LRMoE and a brief comparison between our work and existing literature are provided in \cref{sec:Mixed-LRMoE-remarks}

From a modelling perspective, \cite{Fung2022MixedLRMoETheory} shows that the Mixed LRMoE is \emph{dense} in the space of any mixed effects models subjected to mild regularity conditions. It means that the Mixed LRMoE is flexible enough to resemble any complex characteristics inherited from any mixed effects models, including the joint distribution, the regression pattern, the random intercept, and the random slope, to an arbitrary degree of accuracy. This theoretical result is an extension of \cite{fung2019class}, whereby the LRMoE is shown to be dense in the space of regression models, justifying the versatility and parsimony of the LRMoE with a reduced model structure. The addition of random effects is also crucial for modelling the temporal dependence between observations across different policy years in a large, multiyear insurance portfolio. These desirable features of Mixed LRMoE make it a powerful tool for \emph{a posteriori} risk classification and ratemaking, as demonstrated by our real data analysis in \cref{sec:RealDataAnalysis}, where we apply our proposed framework to an automobile insurance dataset. Our model is shown to outperform classical models in terms of goodness-of-fit to data, while offering fair and interpretable risk classification and ratemaking which accurately reflect policyholders’ claim history.

Besides methodologically applying the Mixed LRMoE model to \emph{a posteriori} risk classification and ratemaking, our second major contribution is the development of a variational inference (VI) algorithm for parameter estimation and posterior inference of random effects. In general, when random effects are included in regression models, parameter estimation and inference may be challenging due to typically intractable likelihood functions. As a classical approach, one may consider applying the Best Linear Unbiased Predictor (BLUP) procedure for obtaining the realization of random effects, combined with Restricted/Residual Maximum Likelihood (REML) for estimating the model parameters, see e.g.~\cite{henderson1973sire}, \cite{henderson1975best}, \cite{mclean1991unified} for Linear Mixed Models, \cite{mcgilchrist1994estimation} and \cite{mcgilchrist1995derivation} for Generalized Linear Mixed Models, and \cite{yau2003finite} and \cite{ng2007extension} for MoE models. Alternatively, one may choose to estimate the parameters from the marginal likelihood by numerically integrating out the random effects using e.g. the Gauss-Hermite Quadrature (\cite{pinheiro1995approximations}) or the Laplace approximation (e.g.~\cite{breslow1993approximate} and \cite{raudenbush2000maximum}). One may also apply Markov Chain Monte Carlo (MCMC) methods (e.g.~\cite{zeger1991generalized}, \cite{booth1999maximizing} and \cite{brooks2011handbook}) for generating samples of random effects from their posterior distribution given the observed data, based on which the posterior of model parameters can also be obtained. A comparison of these methods for models with random effects can be found in \cite{browne2006comparison}. However, the aforementioned methods may not be suitable for the application of \emph{a posteriori} risk classification and ratemaking. For example, when working with large insurance portfolios, it desirable to develop an algorithm which scales with the number of random effects and the size of datasets, which may be difficult for numerical integration or MCMC methods. Also, it is desirable to obtain posteriori distributions, rather than point estimates, of certain quantities of interest (e.g.~\emph{a posteriori} premium based on different premium principles), which are not produced by either BLUP or numerical integration methods. Hence, in place of these classical methods, we opt to use VI primarily for its superior speed and scalability for large insurance portfolios. Besides estimating model parameters with computational efficiency, our VI algorithm also directly produces the approximated posterior distribution of random effects for each individual policyholder, which is key for \emph{a posteriori} risk classification and ratemaking for future policy years. Further, while VI methods have been widely used in the machine learning community as an alternative to computationally more expensive methods such as MCMC (\cite{blei2017variational}), there has been little application of VI in the actuarial literature (see e.g.~\cite{kuo2020individual} and \cite{gomes2021insurance}). We hope our paper serves as another example to showcase the potentials of VI methods for analyzing the ever-growing amount of data available for insurance applications.

The remainder of this paper is organized as follows. \cref{sec:ModelFramework} reviews the LRMoE model and introduces the Mixed LRMoE. Then, \cref{sec:ParameterEstimation} develops a stochastic variational Expectation-Conditional-Maximization (ECM) algorithm for estimating model parameters and inferring the posterior distribution of random effects. Next, \cref{sec:SimulationStudies} presents two simulation studies which aim to numerically illustrate and examine the proposed estimation algorithm, and \cref{sec:RealDataAnalysis} contains an application of our proposed framework on a real insurance dataset. Finally, \cref{sec:Conclusion} concludes with a brief discussion and outlook for future research directions.

\section{Modelling Framework} \label{sec:ModelFramework}

In this section, we first give an overview of the LRMoE modelling framework, including model formulation, theoretical properties, implementation and application in actuarial contexts. Then, we extend the LRMoE model with random effects to account for the temporal dependence across different policy years. Finally, we provide some discussion on the Mixed LRMoE and a brief comparison between our work and existing literature.

\subsection{Overview of LRMoE} \label{sec:LRMoE-overview}

The LRMoE model first introduced in \cite{fung2019class} is formulated as follows. Let $\bm{x}_i$ denote a $P$-dimensional vector of covariates of policyholder $i$ such as demographic information and vehicle specification. Given $\bm{x}_i$, the policyholder is classified into one of $g$ latent risk classes by the logit \emph{gating function}
\begin{equation}
    \pi_{j}(\bm{x}_{i}; \bm{\alpha}) = \frac{\exp(\bm{\alpha}^{T}_{j}\bm{x}_{i})}{\sum_{j^{\prime}=1}^{g} \exp(\bm{\alpha}^{T}_{j^{\prime}}\bm{x}_{i}) }, \quad j = 1, 2, \dots, g,
\end{equation}
where $\bm{\alpha}_{j}$ is a vector of regression coefficients for latent class $j$. Within each latent class $j$, a $D$-dimensional vector of response variable(s) $\bm{y}_i$ such as claim frequency and severity is modelled by an \emph{expert function} $f_{j}(\bm{y}_i; \bm{\psi}_j)$, where $\bm{\psi}_j$ denotes the parameters of the expert function. Consequently, the likelihood function for a portfolio of $n$ policyholders is given by
\begin{equation} \label{eq:LRMoE-likelihood}
    L(\bm{\alpha}, \bm{\Psi}; \bm{X}, \bm{Y}) = \prod_{i=1}^{n}   \left[ \sum_{j=1}^{g}  \pi_{j}(\bm{x}_{i}; \bm{\alpha}) f_{j}(\bm{y}_i; \bm{\psi}_j) \right]
\end{equation}
where $\bm{\alpha} = (\bm{\alpha}_{1}^T, \bm{\alpha}_{2}^T, \dots, \bm{\alpha}_{g}^T)^T$ and $\bm{\Psi} = \{ \bm{\psi}_1, \bm{\psi}_2, \dots, \bm{\psi}_g \}$ are the model parameters to estimate given the observed data $(\bm{X}, \bm{Y}) = \{ (\bm{x}_i, \bm{y}_i): i=1, 2, \dots, n \}$. We assume conditional independence among all dimensions in $\bm{y_i}$ given the latent class $j$ such that $f_{j}(\bm{y}_i; \bm{\psi}_j) = \prod_{d=1}^{D}f_{jd}(y_{id}; \bm{\psi}_{jd})$ for $d=1, 2, \dots, D$, where $y_{id}$ is the $d$-th dimension in $\bm{y}_i$ and $f_{jd}$ is the expert function for $y_{id}$ with parameters $\bm{\psi}_{jd}$.

The LRMoE model can be viewed as a simplification of the general MoE model (see e.g.~\cite{jordan1994hierarchical}), whereby the gating function is restricted to multiple logistic functions and the regression on covariates in the expert functions is eliminated. It is shown in \cite{fung2019class} that such simplification will not reduce modelling flexibility, provided the expert functions satisfy some mild conditions. In other words, the LRMoE model is capable of achieving the same level of goodness-of-fit as the general mixture-of experts with a much simpler model structure. In the meantime, the simplified model structure of LRMoE provides the following intuitive model interpretation in insurance contexts. Based on covariates $\bm{x}_i$ which are indicative of individual risk profiles, policyholders are classified into latent risk groups by a commonly used function for classification problems. Within the same latent group $j$, the individual risk profiles are naturally assumed to be homogeneous by sharing the same expert function $f_{j}(\bm{y}_i; \bm{\psi}_j)$ whose parameters are independent of policyholder information.

Thanks to its flexibility and interpretability, the LRMoE model has been applied to many actuarial modelling problems. \cite{fung2019classapplication} used it for modelling correlated claim frequencies of two types of automobile insurance coverage, where the LRMoE mixture of Erlang Count experts is shown to outperform the negative binomial GLM (with and without zero inflation). \cite{fung2020fittingcensor} discussed fitting LRMoE to censored and truncated data which are commonly encountered when modelling claim severity or reporting delays. The extended model is applied to insurance pricing with policy deductibles and prediction of incurred but not reported (IBNR) claims. In \cite{fung2021mixture}, the LRMoE is further extended to include composite or slicing expert functions which account for multi-modal and heavy-tailed distributions. For implementation of LRMoE, software packages written in R (\cite{tseung2020lrmoeR}) and in Julia (\cite{tseung2021lrmoejl}) are readily available for use, which offer a wide selection of expert functions commonly used for actuarial modelling and utility functions for predictive analysis and model visualization.

As with many mixture models, parameter estimation for LRMoE is done using the Expectation-Conditional-Maximization (ECM) algorithm (see e.g.~\cite{DEMPSTER1977EM} and \cite{Mclachlan2004Finite}). Details of the ECM algorithm for LRMoE can be found in the papers cited above. For Mixed LRMoE, we combine the same ECM algorithm with VI methods in order to deal with intractable marginal likelihood due to the presence of random effects, which will be presented in \cref{sec:ParameterEstimation}.

\subsection{Formulation of Mixed LRMoE} \label{sec:Mixed-LRMoE-overview}

\afterpage{
\begin{figure}[!ht]
    \centering
    \includegraphics[width=0.9\textwidth]{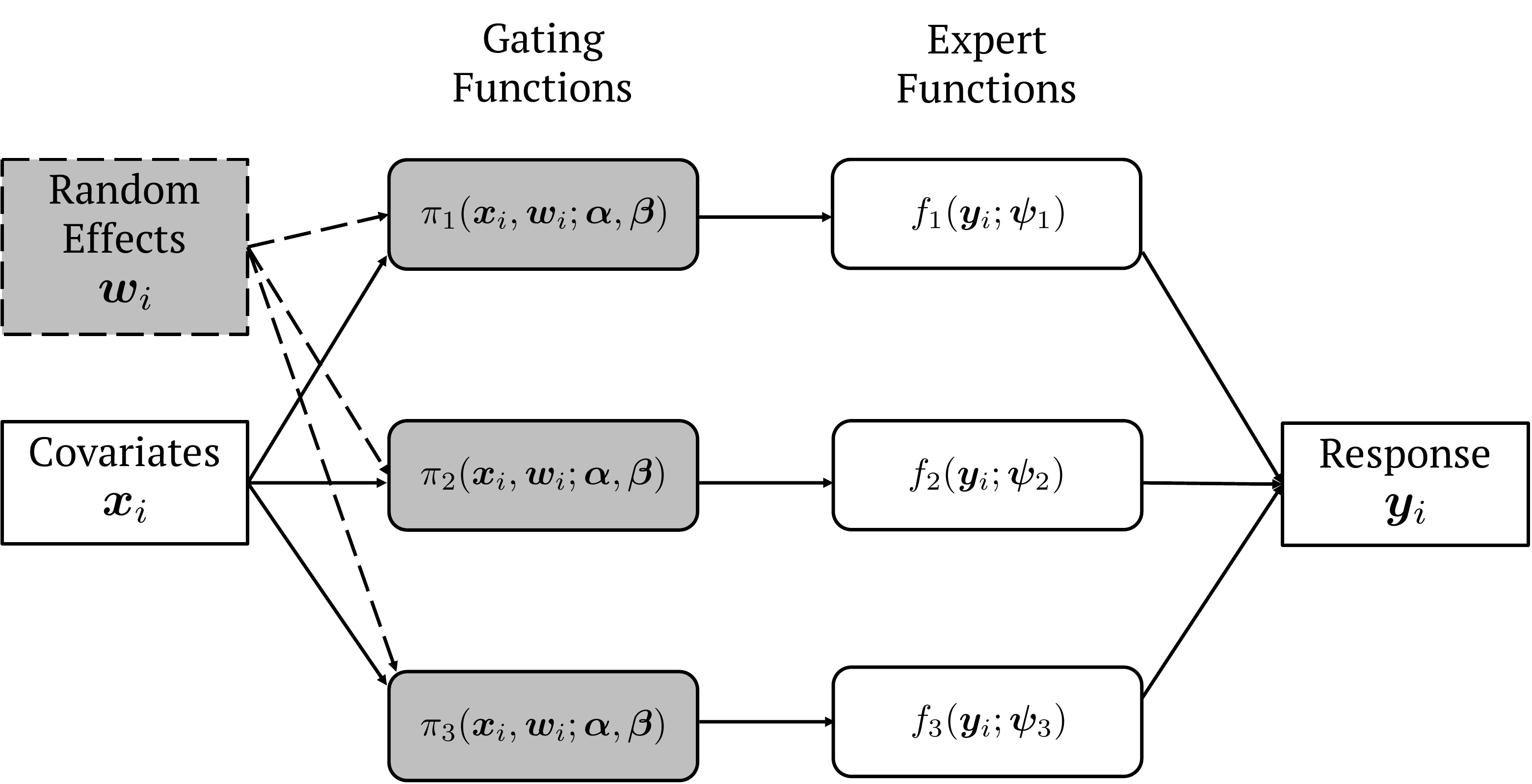}
    \caption{Model structure of a 3-class Mixed LRMoE model. The shaded boxes indicate the addition of random effects to the original LRMoE model in order to account for policyholder-level individual risks and temporal dependence among different policy years for the same policyholder.}
    \label{fig:mixed-LRMoE-structure}
\end{figure}

\vspace{1em}
% \vfill
% \pagebreak
}

In the context of \emph{a posteriori} risk classification and ratemaking, it is important to utilize information about policyholders' claim history to make predictions for the upcoming policy years. In effect, one takes advantage of the dependence structure in the claim history across different policy years generated by the same policyholder. Note that such dependence structure has not been accounted for by the LRMoE model, due to the assumption of independence between observations $(\bm{x}_i, \bm{y}_i)$ as indicated by the likelihood function in \cref{eq:LRMoE-likelihood}. To incorporate dependence between observations across different policy years, we propose to add policyholder-level random effects to the LRMoE model, which results in the Mixed LRMoE model. In this subsection, we first formulate the Mixed LRMoE in a general setting following \cite{Fung2022MixedLRMoETheory}, and then discuss the special case with only policyholder-level random effects.

Assume each observation $(\bm{x}_i, \bm{y}_i)$ is equipped with a vector of random effects $\bm{w}_i = (w_{i1}, w_{i2}, \dots, w_{iL})$, where $L$ is the total number of levels of different random effects. For the $l$-th level of random effect, $l=1, 2, \dots, L$, we assume there are in total $S_l$ factors $\{ w_{l}^{(s)}\}_{s=1, 2, \dots, S_l}$, and each observation $i$ is mapped into one of these factors by a known function $c_l(\cdot)$ such that $w_{il} = w_{i'l} = w_{l}^{(s)}$ if $c_l(i) = c_l(i') = s$ for $s = 1, 2, \dots, S_l$. Equivalently, the mapping function $c_l(i)$ can be represented by a $S_l$-vector $\bm{t}_{il}$ where exactly the $c_l(i)$-th element is one and the others are zero (see also \cref{fig:mixed-LRMoE-multiyear-example} for an example).

Let $\bm{w} = \{w_{l}^{(s)}\}_{l=1, 2, \dots, L; s=1, 2, \dots, S_l}$ denote the collection of random effects across all levels and all factors, which are assumed to be independent across $l$ and $s$.  We also assume their distribution and density functions are pre-specified by $\bm{\Phi}(\cdot)$ and $\bm{\phi}(\cdot)$ with no extra parameters such that
\begin{equation} \label{eq:disn_random_effect}
    \bm{\Phi}(\bm{w}) = \prod_{l=1}^{L}\prod_{s=1}^{S_l} \Phi_l(w_l^{(s)}) \quad \textrm{and} \quad \bm{\phi}(\bm{w}) = \prod_{l=1}^{L}\prod_{s=1}^{S_l} \phi_l(w_l^{(s)}).
\end{equation}
where $\Phi_l(\cdot)$ and $\phi_l(\cdot)$ are, respectively, the distribution and density functions for the $l$-th level of random effects $\{ w_{l}^{(s)}\}_{s=1, 2, \dots, S_l}$ for $l=1, 2, \dots, L$. In general, one may specify \emph{a priori} any distribution for $\bm{\Phi}(\cdot)$, but a common choice for random effects is the normal distribution. In this paper, we will set each $\Phi_l(\cdot)$ to be a standard normal distribution for $l=1, 2, \dots, L$. More discussions on the choice of $\bm{\Phi}(\cdot)$ are given in \cref{sec:Mixed-LRMoE-dense}.

Similar to the covariates $\bm{x}_i$, we assume the random effects $\bm{w}_i$ influences only the gating function. In addition, we assume there are coefficients $\bm{\beta}_j$, $j=1, 2, \dots, g$, multiplied to the random effects, which serve as scaling factors that also affect the gating functions and add to the modelling flexibility by compensating the lack of parameters in $\bm{\Phi}(\cdot)$. Consequently, the gating function in a Mixed LRMoE model is given by
\begin{equation}
    \pi_{j}(\bm{x}_{i}, \bm{w}_i; \bm{\alpha}, \bm{\beta}) = \frac{\exp(\bm{\alpha}^{T}_{j}\bm{x}_{i} + \bm{\beta}^{T}_{j}\bm{w}_{i})}{\sum_{j^{\prime}=1}^{g} \exp(\bm{\alpha}^{T}_{j^{\prime}}\bm{x}_{i} + \bm{\beta}^{T}_{j^{\prime}}\bm{w}_{i}) }, \quad j = 1, 2, \dots, g.
\end{equation}

Unlike the gating functions, the expert functions are assumed to be independent of both the covariates $\bm{x}_i$ and the random effects $\bm{w}_i$, as illustrated in \cref{fig:mixed-LRMoE-structure}. Note this is the same assumption used in the LRMoE model without random effects. Consequently, given the realization of random effects $\bm{w}$, the likelihood function of Mixed LRMoE is
\begin{equation}
    \tilde{L}(\bm{\alpha}, \bm{\beta}, \bm{\Psi}; \bm{X}, \bm{Y}, \bm{w}) = \prod_{i=1}^{n}   \left[ \sum_{j=1}^{g}  \pi_{j}(\bm{x}_{i}, \bm{w}_i; \bm{\alpha}, \bm{\beta}) f_{j}(\bm{y}_i; \bm{\psi}_j) \right],
\end{equation}
while the likelihood with random effects integrated out is given by
\begin{equation} \label{eq:marginal-likelihood-mixed-LRMoE}
    L(\bm{\alpha}, \bm{\beta}, \bm{\Psi}; \bm{X}, \bm{Y}) =  \bigintsss \tilde{L}(\bm{\alpha}, \bm{\beta}, \bm{\Psi}; \bm{X}, \bm{Y}, \bm{w}) \times \bm{\phi}(\bm{w}) d \bm{w} = \E_{\bm{w} \sim \bm{\phi}(\cdot)} \left[ \tilde{L}(\bm{\alpha}, \bm{\beta}, \bm{\Psi}; \bm{X}, \bm{Y}, \bm{w}) \right]
\end{equation}
where $d\bm{w} = \prod_{l=1}^{L}\prod_{s=1}^{S_l} d w_l^{(s)}$ and the subscript of the expectation operator $\E$ indicates the expectation is calculated by integrating out $\bm{w}$ with respect to $\bm{\phi}(\cdot)$.

\subsection{Denseness property of the Mixed LRMoE} \label{sec:Mixed-LRMoE-dense}

The most important property of the Mixed LRMoE is the \emph{denseness} property, which justifies the flexibility of the proposed model in capturing a broad range of complex multilevel data characteristics. While the theoretical result has been rigorously developed by \cite{Fung2022MixedLRMoETheory}, we hereby briefly describe and interpret the result without extensive mathematical treatments.

Let $F(\bm{Y};\bm{\alpha},\bm{\beta},\bm{\Psi}|\bm{X})$ be the joint distribution function of $\bm{Y}$ given $\bm{X}$ under the proposed Mixed LRMoE model, which is given by

\begin{equation}
F(\bm{Y};\bm{\alpha},\bm{\beta},\bm{\Psi}|\bm{X})=
\bigintsss \prod_{i=1}^{n}   \left[ \sum_{j=1}^{g}  \pi_{j}(\bm{x}_{i}, \bm{w}_i; \bm{\alpha}, \bm{\beta}) F_{j}(\bm{y}_i; \bm{\psi}_j) \right] \times \bm{\phi}(\bm{w}) d \bm{w},
\end{equation}
where $F_{j}(\bm{y}_i; \bm{\psi}_j)$ is the distribution function of $f_{j}(\bm{y}_i; \bm{\psi}_j)$. Also, denote $H(\bm{Y}|\bm{X})$ as the joint distribution of $\bm{Y}$ given $\bm{X}$ under an arbitrary mixed effects model. Under some mild regularity conditions, \cite{Fung2022MixedLRMoETheory} proves that for any target mixed effects model $H(\bm{Y}|\bm{X})$, there exists a sequence of model parameters $\{(\bm{\alpha}^{[s]},\bm{\beta}^{[s]},\bm{\Psi}^{[s]})\}_{s=1,2,\ldots}$ (note that the number of latent risk classes $g$ may increase as $s$ increases) such that $F(\bm{Y};\bm{\alpha}^{[s]},\bm{\beta}^{[s]},\bm{\Psi}^{[s]}|\bm{X})$ converges in distribution to $H(\bm{Y}|\bm{X})$ uniformly on $\bm{X}$ as $s\rightarrow\infty$. Note that the target mixed effects model $H(\bm{Y}|\bm{X})$ may carry very complicated model characteristics, including but not limited to the joint loss distribution (e.g., distributional multimodality and dependence across business lines), the regression link (e.g., non-linear or interactive influence of policyholder attributes to the losses), the random intercept (e.g., latent impacts to each policyholder), and the random slope (e.g., random effects interact with policyholder attributes). As a result, the \emph{denseness} theorem justifies the versatility of the proposed Mixed LRMoE in simultaneously capturing all these features to an arbitrary degree of accuracy. Moreover, the \emph{denseness} theorem only requires that $\bm{\Phi}(\cdot)$ is continuous. Hence, one has the freedom to choose any continuous distributions for the random effects without impeding the flexibility of the Mixed LRMoE. Motivated by the computational convenience (see \cref{sec:ParameterEstimation} below), we select $\Phi_l(\cdot)$ (\cref{eq:disn_random_effect}) to be a standard normal distribution, such that $\bm{\Phi}(\cdot)$ follows a multivariate standard normal distribution.

\subsection{Remarks on Mixed LRMoE} \label{sec:Mixed-LRMoE-remarks}

\afterpage{
\begin{figure}[!ht]
    \centering
    \includegraphics[width=0.8\textwidth]{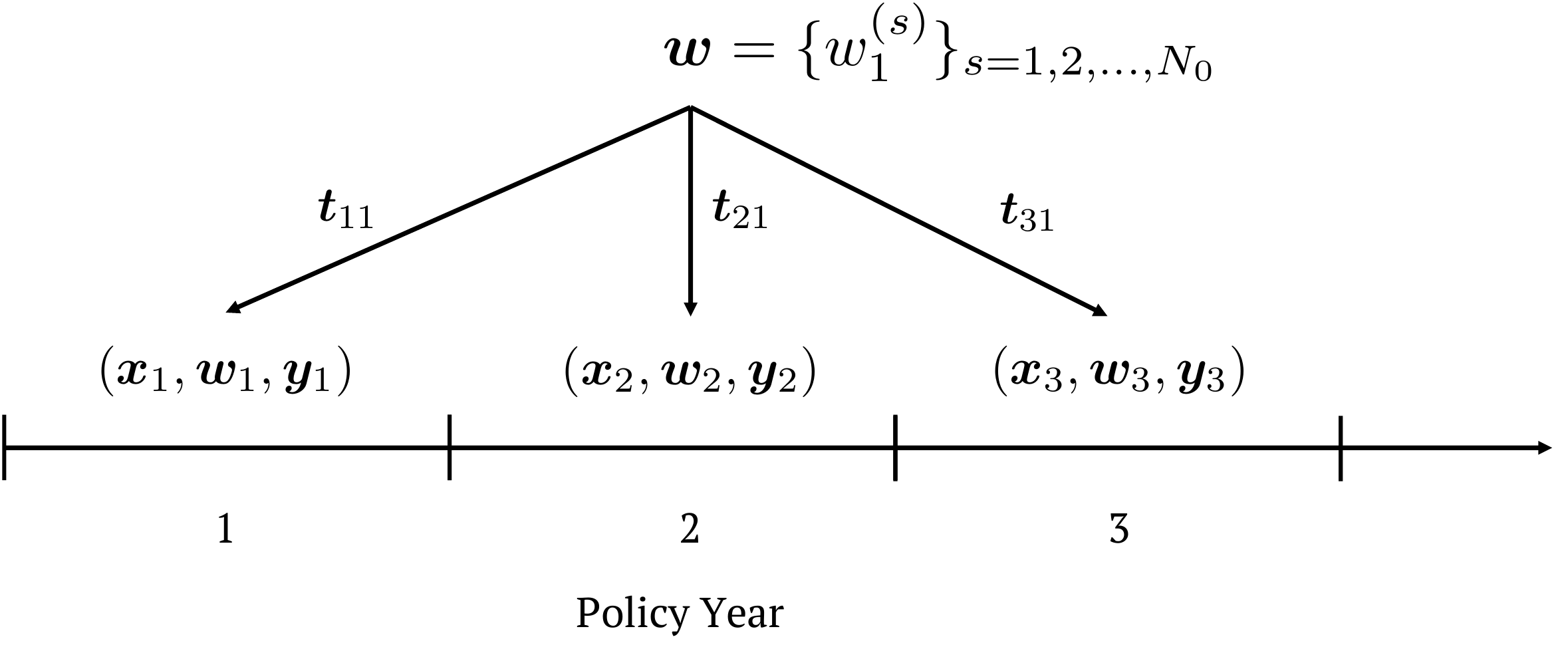}
    \caption{Example of a Mixed LRMoE model with $L=1$ level of random effects on $N_0$ unique policyholders. A policyholder with three years of claim history is represented by three separate yet dependent observations in the dataset, where the dependence is modelled by sharing the same factor in the random effect $\bm{w}$. Assuming this individual is encoded as the first factor $w_{1}^{(1)}$, the mapping vectors are $\bm{t}_{11} = \bm{t}_{21} = \bm{t}_{31} =  (1, 0, 0, \dots, 0)$ so that $\bm{w}_1 = \bm{w}_2 = \bm{w}_3 = (w_{1}^{(1)})$ are the same 1-length vector of random effect. This can be equivalently described by the mapping function $c_1(1) = c_1(2) = c_1(3) = 1$. Note that some elements in the covariates $\bm{x}_i$ may change over time, such as the policyholder's age.}
    \label{fig:mixed-LRMoE-multiyear-example}
\end{figure}
\vspace{1em}
}

Before proceeding to parameter estimation, we make the following remarks on the model formulation of Mixed LRMoE and provide a brief comparison with existing literature.

Firstly, in \cref{sec:Mixed-LRMoE-overview} we have given a general formulation of Mixed LRMoE with potentially multiple levels of random effects when $L>1$. For the application in \emph{a posteriori} risk classification and ratemaking in this paper, we set $L=1$ to add only policyholder-level random effects. In this case, $n$ is the total number of policy year observations out of $N_0$ unique policyholders, such that each factor in $\{w_{1}^{(s)}\}_{s=1, 2, \dots, N_0}$ represents the individual risk of one unique policyholder. An illustration for one such policyholder is shown in \cref{fig:mixed-LRMoE-multiyear-example}. Other than \emph{a posteriori} risk classification and ratemaking, one may consider applying the Mixed LRMoE to other modelling problems with multiple levels of latent risks, such as modelling geographical risks with a nested structure for random effects with $L=2$ levels, where $l=1$ represents city-level random effects and $l=2$ represents the latent risks for specific neighbourhoods. For illustration purposes, we will leave the application of Mixed LRMoE with $L>1$ for future investigation, and only demonstrate a simulation study for $L=2$ in \cref{sec:SimulationStudies}.

Secondly, similar to many previous works such as those cited in \cref{sec:Introduction}, our paper also utilizes random effects for modelling temporal dependence among different policy years of the same policyholder, but we have done so in a slightly different fashion. Many previous papers have proposed mixed models whereby the certain model parameters are shared across different observations. For example, one may assume the claim frequency $N_{it}$ of policyholder $i$ in the $t$-th year follows $\textrm{Poisson}(\theta_{it})$, and then uses the observed data $\{N_{it}: t=1, 2, \dots\}$ to infer the posterior of the intensity parameter. In contrast, our formulation of the Mixed LRMoE treats the random effects $\bm{w}$ in a similar way as the fixed effects $\bm{x}_i$, which essentially serve as a regressor in the gating function. Rather than imposing certain changing dynamics on model parameters, the formulation of Mixed LRMoE actually resembles, to a large extent, classical approaches of longitudinal data modelling with random effects, see e.g.~\cite{diggle2002analysis} and \cite{fitzmaurice2012applied}.

Finally, the Mixed LRMoE model shares varying degrees of similarity with previous works which attempt to incorporate random effects in the general MoE framework. For example, \cite{yau2003finite} proposes a two-component MoE with random effects in both the logit gating function and normal experts. \cite{ng2007extension} considers a similar framework but uses Bernoulli experts for a classification problem, while \cite{ng2014mixture} adds random effects only to the expert functions. In contrast, our present work focuses on a specific subclass of Mixed MoE model where random effects only influence the latent class probabilities through the gating function, while the expert functions are kept independent of covariates and random effects. Besides possessing the same level of modelling flexibility due to denseness (as discussed in \cref{sec:Mixed-LRMoE-dense}), this simplified model structure leads to an easier implementation of parameter estimation. As will be evident in \cref{sec:ParameterEstimation}, since the estimation procedures of gating and expert functions can be separated to some extent, the Mixed LRMoE model actually allows for more flexible choices and combinations of expert functions which are customized to different modelling problems (see also \cref{sec:Conclusion}). By restricting the random effects to only the gating functions, we are able to develop a unified estimation algorithm which caters for different choices and combinations of expert functions.

\section{Parameter Estimation} \label{sec:ParameterEstimation}

In this section, we develop a stochastic variational ECM algorithm for estimating model parameters and for inferring the posterior distribution of random effects for Mixed LRMoE. We first present an overview of variational inference methods in general, and then provide details of the implementation for Mixed LRMoE with one single type of random effect. Discussion on model identifiability, model selection and generalization of this algorithm is given at the end of this section.

\subsection{Overview of Variational Inference}

In this subsection, we first provide an overview and motivation of variational inference methods. We start with the exact posterior distribution of random effects $\bm{w}$
\begin{equation} \label{eq:mixed-LRMoE-exact-posterior}
    \bm{p}(\bm{w}; \bm{\alpha}, \bm{\beta}, \bm{\Psi} | \bm{X}, \bm{Y}) \propto \prod_{i=1}^{n}   \left[ \sum_{j=1}^{g}  \pi_{j}(\bm{x}_{i}, \bm{w}_i; \bm{\alpha}, \bm{\beta}) f_{j}(\bm{y}_i; \bm{\psi}_j) \right] \times \bm{\phi}(\bm{w})
\end{equation}
which may be complicated due to the dependence on both the model parameters $(\bm{\alpha}, \bm{\beta}, \bm{\Psi})$ and the observed data $(\bm{X}, \bm{Y})$. To circumvent this numerical challenge, we assume the exact posterior can be reasonably approximated by a \emph{variational distribution} $\bm{q}(\bm{w} ; \bm{\Theta})$ where $\bm{\Theta}$ is the \emph{variational parameters}, which are assumed to be independent of the model parameters and observed data. This produces a numerically more tractable lower bound of the marginal likelihood in \cref{eq:marginal-likelihood-mixed-LRMoE}, also known as the Evidence Lower Bound (ELBO) in the variational inference literature. More specifically, by taking logarithm of \cref{eq:marginal-likelihood-mixed-LRMoE}, utilizing the variational distribution, and applying Jensen's inequality, we obtain the following ELBO of the marginal loglikelihood.
\begin{equation}
\begin{aligned} \label{eq:Mixed-LRMoE-ELBO-derivation}
    \ell(\bm{\alpha}, \bm{\beta}, \bm{\Psi}; \bm{X}, \bm{Y}) &= \log \bigintsss \tilde{L}(\bm{\alpha}, \bm{\beta}, \bm{\Psi}; \bm{X}, \bm{Y}, \bm{w}) \times \bm{\phi}(\bm{w}) d \bm{w} \\
    &= \log \bigintsss \tilde{L}(\bm{\alpha}, \bm{\beta}, \bm{\Psi}; \bm{X}, \bm{Y}, \bm{w}) \times \frac{\bm{\phi}(\bm{w})}{\bm{q}(\bm{w};\bm{\Theta})} \times \bm{q}(\bm{w};\bm{\Theta}) d \bm{w} \\
    &\geq \E_{\bm{w} \sim \bm{q}(\cdot;\bm{\Theta})} \left[ \log\tilde{L}(\bm{\alpha}, \bm{\beta}, \bm{\Psi}; \bm{X}, \bm{Y}, \bm{w}) + \log\bm{\phi}(\bm{w}) - \log\bm{q}(\bm{w};\bm{\Theta}) \right]  \\
    &= \E_{\bm{w} \sim \bm{q}(\cdot;\bm{\Theta})} \left[ \log\tilde{L}(\bm{\alpha}, \bm{\beta}, \bm{\Psi}; \bm{X}, \bm{Y}, \bm{w}) \right] - \textrm{KL}\left[ \bm{q}(\bm{w} ; \bm{\Theta}) || \bm{\phi}(\bm{w}) \right] \\
    &:= \underline{\ell}(\bm{\alpha}, \bm{\beta}, \bm{\Psi}, \bm{\Theta}; \bm{X}, \bm{Y})
\end{aligned}
\end{equation}
where $\textrm{KL}\left[ \bm{q}(\bm{w} ; \bm{\Theta}) || \bm{\phi}(\bm{w}) \right]$ is the Kullback-Leibler (KL) divergence between the variational posterior $\bm{q}(\bm{w} ; \bm{\Theta})$ and the prior $\bm{\phi}(\bm{w})$ of random effects.

Instead of directly maximizing the marginal likelihood in \cref{eq:marginal-likelihood-mixed-LRMoE}, we aim to maximize the ELBO $\underline{\ell}(\bm{\alpha}, \bm{\beta}, \bm{\Psi}, \bm{\Theta}; \bm{X}, \bm{Y})$ in \cref{eq:Mixed-LRMoE-ELBO-derivation}, hoping that the optimal parameters which maximize this lower bound are close to the true optimal parameters which maximize the actual loglikelihood. The main advantage is the tractability of the approximate posterior of random effects $\bm{w}$, which is essentially specified by parameters $\bm{\Theta}$ independent of all the other model parameters and observed data. As will be evident in the next subsection, sampling from the approximated posterior is easier and faster than MCMC methods, since the latter works with a more complex exact posterior and typically requires a burn-in period. This may offer significant numerical efficiency, especially in high-dimensional cases where there are many types of random effects and each type of random effect has many levels. Meanwhile, the obvious trade-off is obtaining only the approximated solutions to the estimated model parameters and the approximated posterior distributions of the random effects. While the goodness of approximation and convergence properties for variational inference remain an open problem (see e.g.~\cite{blei2017variational}), our numerical simulations in \cref{sec:SimulationStudies} and real data analysis in \cref{sec:RealDataAnalysis} show promising results.  This may serve as an empirical evidence for applying variational inference methods to insurance problems where an approximated solution may be acceptable in the presence of large datasets.

For variational inference, one needs to specify a family of parametric distributions for the approximated posterior $\bm{q}(\bm{w} ; \bm{\Theta})$. In this paper, we follow standard practices and use the \emph{mean-field variational family}, whereby the posterior of latent variables, i.e.~random effects $\bm{w}$, is a factorized multivariate normal distribution. More specifically, we assume the posterior of $w_{l}^{(s)}$ is a normal distribution with mean $\mu_{l}^{(s)}$ and standard deviation $\sigma_{l}^{(s)}$ for $s=1, 2, \dots, S_l$ and $l=1, 2, \dots, L$, which are independent across all levels $l$ and all factors $s$. Mathematically,
\begin{equation}
    \bm{q}(\bm{w} ; \bm{\Theta}) = \prod_{l=1}^{L}\prod_{s=1}^{S_l} \frac{1}{\sqrt{2\pi(\sigma_{l}^{(s)})^2}}\exp\left[ -\frac{1}{2(\sigma_{l}^{(s)})^2} (w_{l}^{(s)} - \mu_{l}^{(s)})^2\right].
\end{equation}

For notational convenience, we write $\bm{\Theta} = \{(\bm{\mu}_l, \bm{\Sigma}_l)\}_{l=1, 2, \dots, L}$, where $\bm{\mu}_l = (\mu_{l}^{(1)}, \mu_{l}^{(2)}, \dots, \mu_{l}^{(S_l)})^T$ is the posterior mean vector and $\bm{\Sigma}_l = \textrm{diag}((\sigma_{l}^{(1)})^2, (\sigma_{l}^{(2)})^2, \dots, (\sigma_{l}^{(S_l)})^2)$ the diagonal covariance matrix for the $l$-th level of random effect.

When $L=1$, given the factorization of likelihood across $s=1, 2, \dots, S_1$, different factors of the same level of random effect are in fact independent, both in the prior and the posterior distribution. Hence, in our application of the Mixed LRMoE with only policyholder-level random effects, the only source of error of variational inference is the approximation of the exact posterior by a normal distribution. However, when there are multiple types of random effects (e.g.~the multilevel example in \cref{sec:Mixed-LRMoE-remarks}), especially in the case of certain dependence structures (e.g.~multiple crossed random effects), the independence assumption in the mean-field variational family may create an additional source of error of approximation.
\subsection{A Stochastic Variational ECM Algorithm}

With the approach of variational inference and the choice of the mean-field variational family $\bm{q}(\bm{w} ; \bm{\Theta})$, we now develop a stochastic variational ECM algorithm for estimating the model parameters $(\bm{\alpha}, \bm{\beta}, \bm{\Psi})$, as well as inferring the posterior of random effects $\bm{w}$ represented by the variational parameters $\bm{\Theta} = \{(\bm{\mu}_l, \bm{\Sigma}_l)\}_{l=1, 2, \dots, L}$.

On a high level, our estimation algorithm proceeds in an iterative manner which seeks to conditionally maximize the ELBO in \cref{eq:Mixed-LRMoE-ELBO-derivation} with respect to one set of parameters while keeping others fixed. Consequently, the algorithm will ultimately arrive at a local optimum for the ELBO of the marginal loglikelihood. First, we initialize the model parameters $({\bm{\alpha}}, {\bm{\beta}}, {\bm{\Psi}})$ using the clusterized method of moments (CMM), similar to e.g.~\cite{gui2018fitting}. Meanwhile, the variational parameters $\bm{\Theta}$ can be initialized such that $\bm{\mu}_l = \bm{0}$ and $\bm{\Sigma}_l = \bm{I}$ for $l=1, 2, \dots, L$ (i.e.~assuming a multivariate standard normal distribution), which is consistent with standard practices in the VI literature. Then, our algorithm iterates through the following steps until convergence.

\textbf{E-Step}: At iteration $t+1$, given the current model parameters $(\bm{\alpha}^{(t)}, \bm{\beta}^{(t)}, \bm{\Psi}^{(t)})$ and variational parameters $\bm{\Theta}^{(t)} = \{(\bm{\mu}_l^{(t)}, \bm{\Sigma}_l^{(t)})\}_{l=1, 2, \dots, L}$, we calculate the expectation of the complete-data ELBO, which results in the objective function $Q^{(t+1)}(\bm{\alpha}, \bm{\beta}, \bm{\Psi}, \bm{\Theta}; \bm{X}, \bm{Y})$.

\textbf{CM-Steps}:
\begin{enumerate}[label=(\roman*)]
    \item Given the current values of the variational parameters $\bm{\Theta}^{(t)}$, we conditionally maximize the objective function in $(\bm{\alpha}^{(t+1)}, \bm{\beta}^{(t+1)}, \bm{\Psi}^{(t+1)})$.
    \item Given the updated $(\bm{\alpha}^{(t+1)}, \bm{\beta}^{(t+1)}, \bm{\Psi}^{(t+1)})$, find the updated variational parameters $\bm{\Theta}^{(t+1)} = \{(\bm{\mu}_l^{(t+1)}, \bm{\Sigma}_l^{(t+1)})\}_{l=1, 2, \dots, L}$ by optimizing the complete-data ELBO.
\end{enumerate}

Next, we describe each of these steps in more detail. In the E-Step, we augment the usual latent variables $\bm{Z} = \{Z_{ij}, i=1, 2, \dots, n$ and $j=1, 2, \dots, g \}$ such that $Z_{ij}=1$ indicates $\bm{y}_i$ is generated by the $j$-th latent class and $Z_{ij}=0$ otherwise. Consequently, the complete-data ELBO is given by
\begin{equation}
\begin{aligned}
    \underline{\ell}^{c}(\bm{\alpha}, \bm{\beta}, \bm{\Psi}, \bm{\Theta}; \bm{X}, \bm{Y}) &= \E_{\bm{w} \sim \bm{q}(\cdot;\bm{\Theta})} \left[ \sum_{i=1}^{n} \sum_{j=1}^{g} Z_{ij} \log \left[ \pi_{j}(\bm{x}_i, \bm{w}_i; \bm{\alpha}, \bm{\beta}) f_{j}(\bm{y}_i; \bm{\psi}_j) \right]  \right] - \textrm{KL}\left[ \bm{q}(\bm{w} ; \bm{\Theta}) || \bm{\phi}(\bm{w}) \right]
\end{aligned}
\end{equation}

We then calculate the expected value of $\bm{Z}$ given the current values of model and variational parameters, which yields the following objective function $Q^{(t+1)}(\bm{\alpha}, \bm{\beta}, \bm{\Psi}, \bm{\Theta}; \bm{X}, \bm{Y})$ to be maximized in the CM-Steps.
\begin{equation} \label{eq:ELBO-complete-objective}
\begin{aligned}
    Q^{(t+1)}(\bm{\alpha}, \bm{\beta}, \bm{\Psi}, \bm{\Theta}; \bm{X}, \bm{Y}) &= \E_{\bm{Z}} \left[ \E_{\bm{w} \sim \bm{q}(\cdot;\bm{\Theta})} \left[ \log\tilde{L}^{c}(\bm{\alpha}, \bm{\beta}, \bm{\Psi}; \bm{X}, \bm{Y}, \bm{w}) \right] \bigg| \bm{X}, \bm{Y}, \bm{\alpha}^{(t)}, \bm{\beta}^{(t)}, \bm{\Psi}^{(t)}, \bm{\Theta}^{(t)} \right] \\ & - \textrm{KL}\left[ \bm{q}(\bm{w} ; \bm{\Theta}) || \bm{\phi}(\bm{w}) \right] \\
    &= \E_{\bm{w} \sim \bm{q}(\cdot;\bm{\Theta})} \left[ \sum_{i=1}^{n} \sum_{j=1}^{g} z_{ij}^{(t)} \log \left[ \pi_{j}(\bm{x}_i, \bm{w}_i; \bm{\alpha}, \bm{\beta}) f_{j}(\bm{y}_i; \bm{\psi}_j) \right] \right] - \textrm{KL}\left[ \bm{q}(\bm{w} ; \bm{\Theta}) || \bm{\phi}(\bm{w}) \right]
\end{aligned}
\end{equation}
where
\begin{equation} \label{eq:EZ-marginal-on-w}
\begin{aligned}
    z_{ij}^{(t)} &= \E \left[ Z_{ij} \bigg| \bm{X}, \bm{Y}, \bm{\alpha}^{(t)}, \bm{\beta}^{(t)}, \bm{\Psi}^{(t)}, \bm{\Theta}^{(t)} \right] = \E_{\bm{w} \sim \bm{q}(\cdot;\bm{\Theta}^{(t)})} \left[ \E \left[ Z_{ij} \bigg| \bm{X}, \bm{Y}, \bm{w} , \bm{\alpha}^{(t)}, \bm{\beta}^{(t)}, \bm{\Psi}^{(t)} \right] \right]
\end{aligned}
\end{equation}
and the change of order of integration is justified by $\sum_{i=1}^{n} \sum_{j=1}^{g} Z_{ij} \log \left[ \pi_{j}(\bm{x}_i, \bm{w}_i; \bm{\alpha}, \bm{\beta}) \right] \leq 0$. Given the realization of random effects $\bm{w}$, the conditional expectation on the right-hand-side of \cref{eq:EZ-marginal-on-w} is evaluated as
\begin{equation} \label{eq:EZ-conditional-on-w}
\begin{aligned}
    \E \left[ Z_{ij} \bigg| \bm{X}, \bm{Y}, \bm{w}, \bm{\alpha}^{(t)}, \bm{\beta}^{(t)}, \bm{\Psi}^{(t)} \right] = \frac{\pi_{j}(\bm{x}_i, \bm{w}_i; \bm{\alpha}^{(t)}, \bm{\beta}^{(t)}) f_{j}(\bm{y}_i; \bm{\psi}_j^{(t)})}{\sum_{j'=1}^{n} \pi_{j'}(\bm{x}_i, \bm{w}_i; \bm{\alpha}^{(t)}, \bm{\beta}^{(t)}) f_{j'}(\bm{y}_i; \bm{\psi}_{j'}^{(t)})}.
\end{aligned}
\end{equation}

Note that the unconditional expectation of $Z_{ij}$ by integrating out $\bm{w}$ admits no closed-form solution. However, the normality assumption on the posterior of $\bm{w}$ allows for the following numerical evaluation through Monte Carlo simulation which entails little computational burden.
\begin{equation} \label{eq:EZ-sampling-with-MC}
    \hat{z}_{ij}^{(t)} = \frac{1}{M} \sum_{m=1}^{M} \E  \left[ Z_{ij} \bigg| \bm{X}, \bm{Y}, \bm{w}^{[m]}, \bm{\alpha}^{(t)}, \bm{\beta}^{(t)}, \bm{\Psi}^{(t)}, \bm{\Theta}^{(t)} \right]
\end{equation}
where $\bm{w}^{[m]}$ denotes the $m$-th sample of random effects generated from the variational distribution $\bm{q}(\cdot;\bm{\Theta}^{(t)})$.

Next, in CM-Step (i), given the current variational parameters $\bm{\Theta}^{(t)}$, the maximization of $Q^{(t+1)}(\bm{\alpha}, \bm{\beta}, \bm{\Psi}, \bm{\Theta}; \bm{X}, \bm{Y})$ is divided into subproblems in $Q_1^{(t+1)}(\bm{\alpha}, \bm{\beta}; \bm{X}, \bm{\Theta}^{(t)})$ and $Q_2^{(t+1)}(\bm{\Psi}; \bm{Y}, \bm{\Theta}^{(t)})$ such that
\begin{equation} \label{eq:Q1-objective-function-alpha-beta}
\begin{aligned}
    Q_1^{(t+1)}(\bm{\alpha}, \bm{\beta}; \bm{X}, \bm{\Theta}^{(t)}) &= \E_{\bm{w} \sim \bm{q}(\cdot;\bm{\Theta}^{(t)})} \left[ \sum_{i=1}^{n} \sum_{j=1}^{g} z_{ij}^{(t)} \log \pi_{j}(\bm{x}_i, \bm{w}_i; \bm{\alpha}, \bm{\beta}) \right]
\end{aligned}
\end{equation}
and
\begin{equation}
\begin{aligned}
    Q_2^{(t+1)}(\bm{\Psi}; \bm{Y}, \bm{\Theta}^{(t)}) &= \E_{\bm{w} \sim \bm{q}(\cdot;\bm{\Theta}^{(t)})} \left[ \sum_{i=1}^{n} \sum_{j=1}^{g} z_{ij}^{(t)} \log f_{j}(\bm{y}_i; \bm{\psi}_j) \right] = \sum_{i=1}^{n} \sum_{j=1}^{g} z_{ij}^{(t)} \log f_{j}(\bm{y}_i; \bm{\psi}_j).
\end{aligned}
\end{equation}

Given the realization of random effects $\bm{w}$, the right-hand-side of \cref{eq:Q1-objective-function-alpha-beta} without the expectation operator can be maximized using the iteratively re-weighted least squares (IRLS) method (see e.g.~\cite{jordan1994hierarchical} and \cite{fung2019classapplication}). To account for the randomness in $\bm{w}$, we adapt the deterministic IRLS procedure to its stochastic version which is described in detailed in \cref{appendix:detailed-estimation-IRLS}. In the meantime, given $z_{ij}^{(t)}$ obtained from the E-Step, the maximization of $Q_2^{(t+1)}(\bm{\Psi}; \bm{Y}, \bm{\Theta}^{(t)})$ over the expert parameters $\bm{\Psi}$ proceeds exactly the same as described in e.g.~\cite{fung2019classapplication} and \cite{tseung2021lrmoejl}, which is independent of the variational parameters $\bm{\Theta}^{(t)}$. Details are omitted here and we refer interested readers to the cited papers.

Finally, in CM-Step (ii), the complete-data ELBO is maximized over the variational parameters $\bm{\Theta} = \{(\bm{\mu}_l, \bm{\Sigma}_l)\}_{l=1, 2, \dots, L}$  given the updated model parameters $(\bm{\alpha}^{(t+1)}, \bm{\beta}^{(t+1)}, \bm{\Psi}^{(t+1)})$. The variational parameters affect the objective function in \cref{eq:ELBO-complete-objective} through both the expectation operator $\E_{\bm{w} \sim \bm{q}(\cdot ;\bm{\Theta})} \left[ \cdot \right]$ and the KL divergence term $\textrm{KL}\left[ \bm{q}(\bm{w} ; \bm{\Theta}) || \bm{\phi}(\bm{w}) \right]$. Assuming the mean-field variational family $\bm{q}(\cdot;\bm{\Theta})$, the optimization over $\bm{\Theta}$ can be effectively done by a standard reparameterization technique on the random effects $\bm{w}$ combined with a simple gradient descent. For brevity, details are deferred to \cref{appendix:detailed-estimation-VI-params}.

 In addition to the estimated model parameters $(\hat{\bm{\alpha}}, \hat{\bm{\beta}}, \hat{\bm{\Psi}})$, our algorithm also yields the variational parameters $\hat{\bm{\Theta}} = \{(\hat{\bm{\mu}}_l, \hat{\bm{\Sigma}}_l)\}_{l=1, 2, \dots, L}$ which completely specify the approximated posterior distribution of random effects $\bm{w}$. Despite no closed-form formulas for various quantities of interests such as the posterior mean of response $\bm{y}_i$ (see also \cref{sec:RealDataAnalysis}), their approximated values can be efficiently calculated by sampling from the variational posterior distribution which is assumed to be multivariate normal.

\subsection{Model Identifiability and Selection} \label{sec:model-identifiability-selection}

As with many mixture models, certain restrictions are imposed for the Mixed LRMoE to be identifiable when conducting parameter estimation. In order to avoid label-switching between latent components (see e.g.~\cite{jiang1999identifiability} and \cite{fung2019classapplication}), we fix $\bm{\alpha}_g = \bm{0}$ and $\bm{\beta}_g = \bm{0}$ as vectors of zeros, so the last latent class serves as a reference class. In addition, we fix $\bm{\beta}_1 = \bm{1}$ as a vector of ones to avoid arbitrary scaling of magnitude and switching of positive and negative signs of the random effects $\bm{w}$. Consequently, we need to estimate the coefficients $\bm{\beta}$ multiplied to the random effects only when there are at least three latent classes (see the examples in \cref{sec:SimulationStudies}).

Model selection when parameters are estimated using variational inference remains an open problem in general. One may accept the ELBO as a good approximation of the marginal likelihood and use it as the basis of model selection, but this has not been justified in theory (\cite{blei2017variational}). Other approaches include sequential selection (\cite{sato2001online}), cross validation (\cite{nott2012regression}) and Generalized Evidence Bounds (\cite{chen2018variational}). For the purpose of this paper, we take a more practical approach by using the standard train-test split and examining the approximated loglikelihood and ELBO on the test set to obtain a conservative gauge of goodness-of-fit. Examples are given in the real data analysis in \cref{sec:RealDataAnalysis}.

\section{Simulation Studies} \label{sec:SimulationStudies}

% Summary tables of simulation studies
\afterpage{
\renewcommand{\arraystretch}{1.5}

\subfile{tables/simulation-1-summary-table}

\vfill

\subfile{tables/simulation-2-summary-table}

\renewcommand{\arraystretch}{1}
\pagebreak
}

% Summary figures of simulation studies
\afterpage{
\begin{figure}[!h]
    \centering
    \caption{Simulation Study I. Left: Fitted vs True random effects. Mid: Fitted vs True probabilities of latent class 1. Right: Histogram of simulated data vs Fitted density.}
    \label{fig:simulation-1-RE}
    \begin{subfigure}{0.32\textwidth}
        \includegraphics[width=\textwidth]{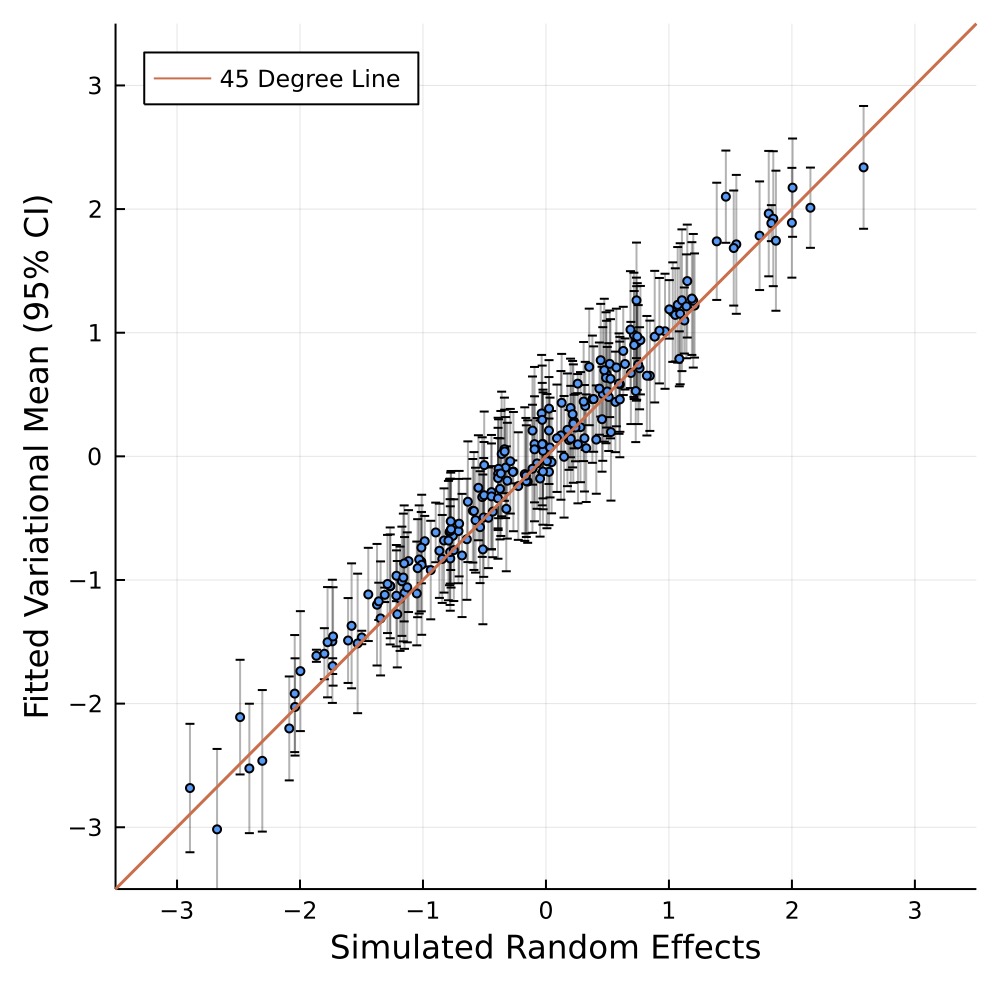}
    \end{subfigure}
    ~
    \begin{subfigure}{0.32\textwidth}
        \includegraphics[width=\textwidth]{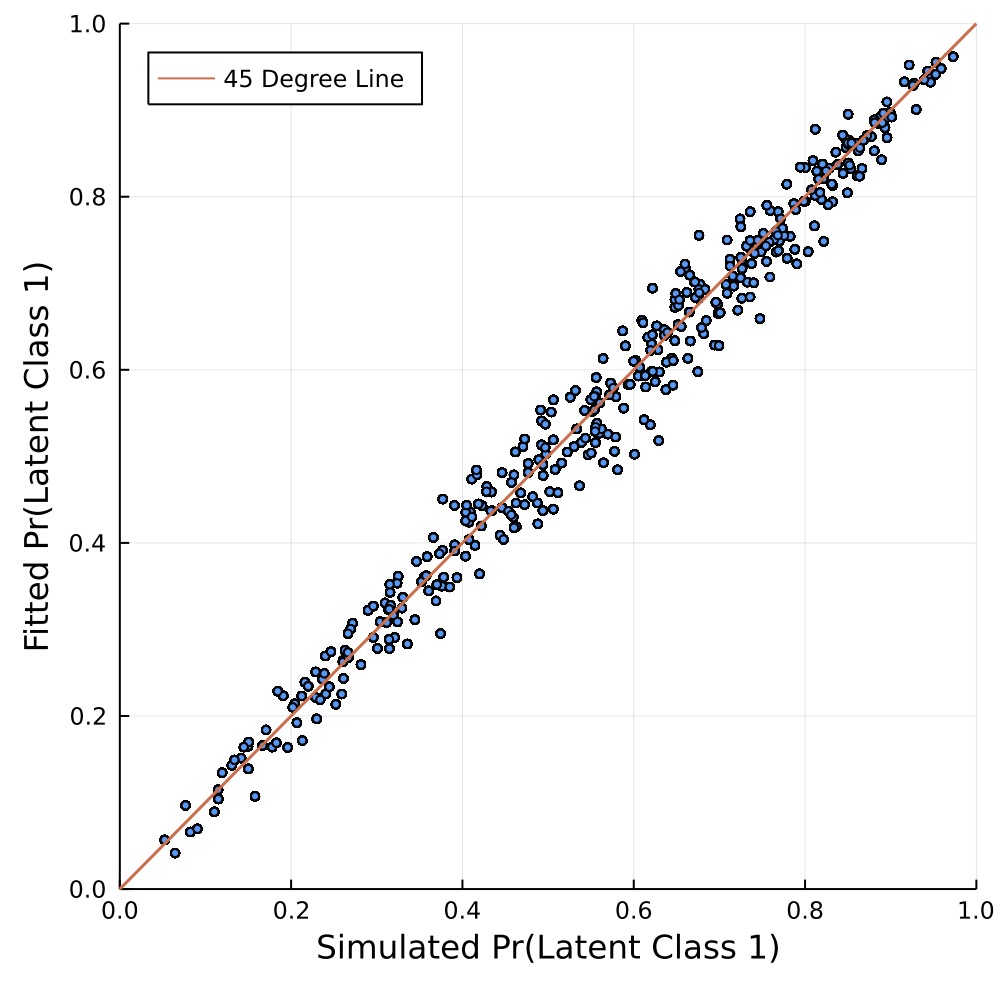}
    \end{subfigure}
    ~
    \begin{subfigure}{0.32\textwidth}
        \includegraphics[width=\textwidth]{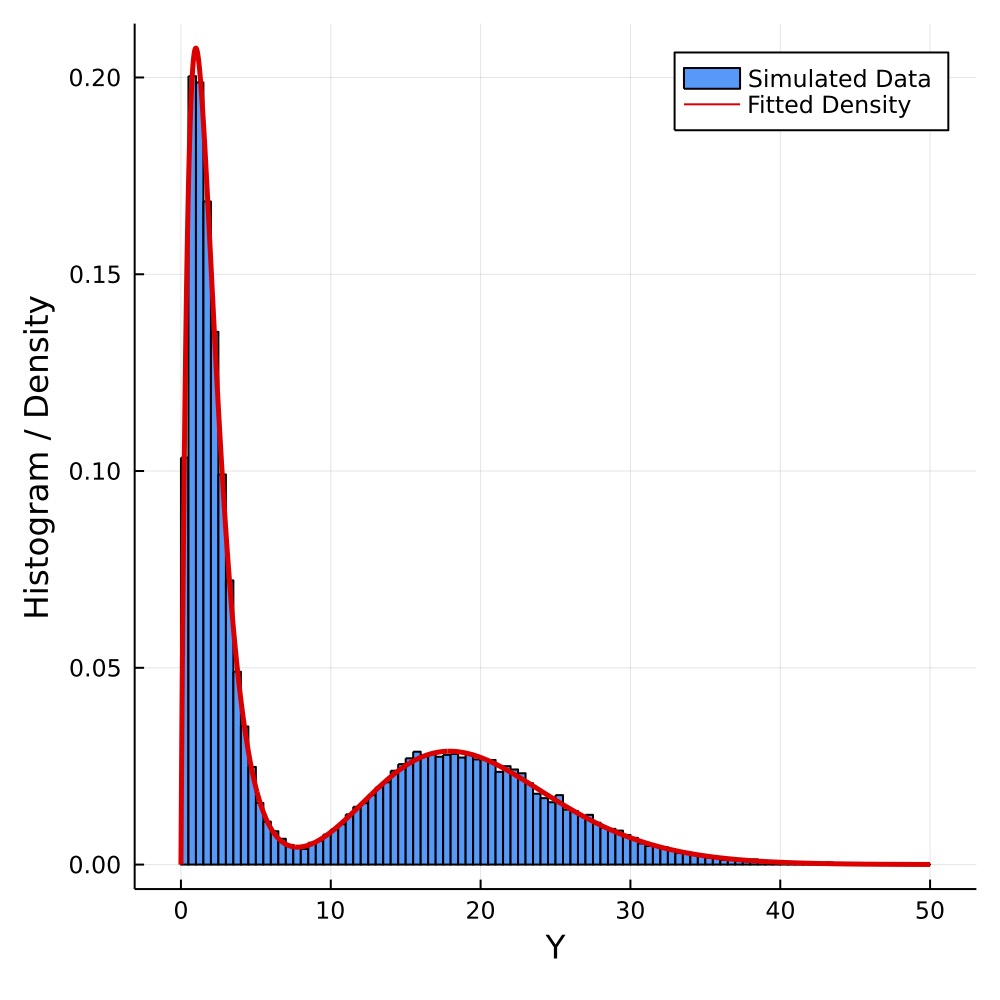}
    \end{subfigure}
\end{figure}

\vfill

\begin{figure}[!h]
    \centering
    \caption{Simulation Study II. Upper Left / Right: Fitted vs True random effects for $l=1$ / $l=2$. Lower Left / Mid: Fitted vs True probabilities of latent class 1 / 2. Lower Right: Histogram of simulated data vs Fitted density.}
    \label{fig:simulation-2-RE}
    \begin{subfigure}{0.32\textwidth}
        \includegraphics[width=\textwidth]{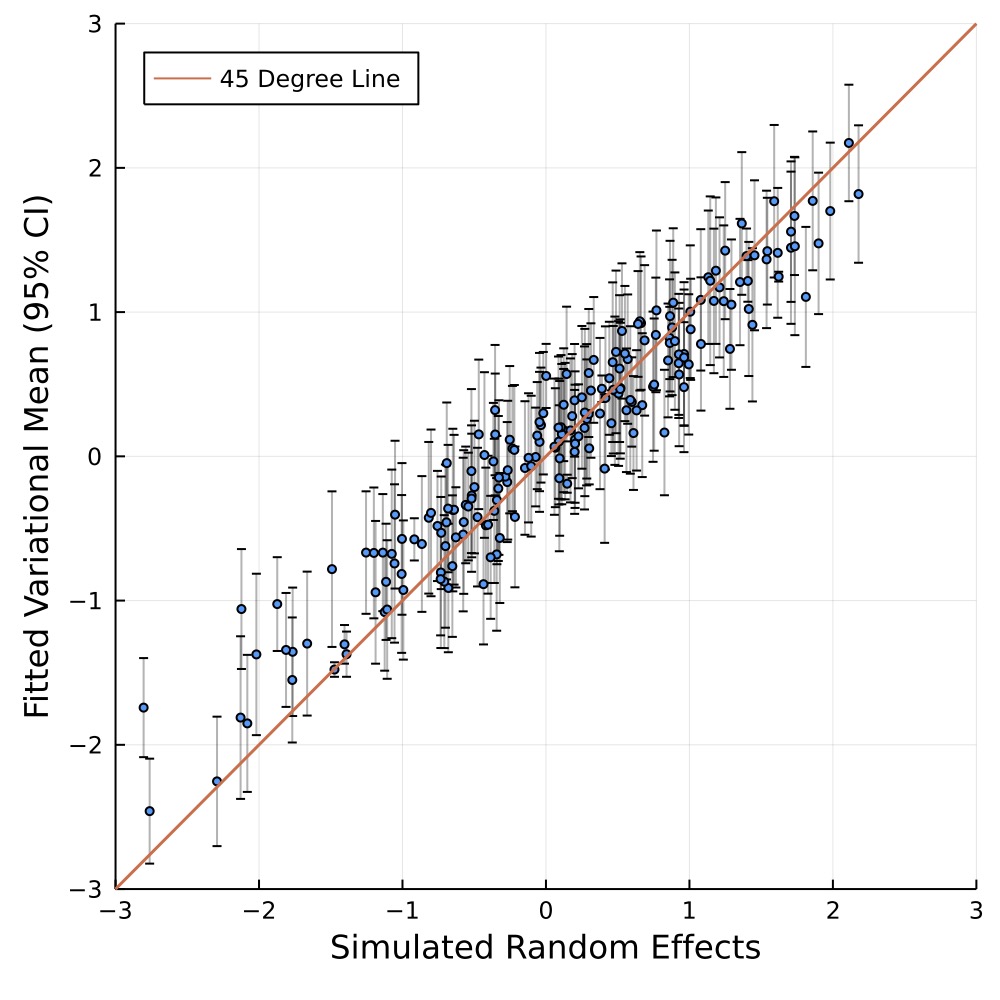}
    \end{subfigure}
    ~
    \begin{subfigure}{0.32\textwidth}
        \includegraphics[width=\textwidth]{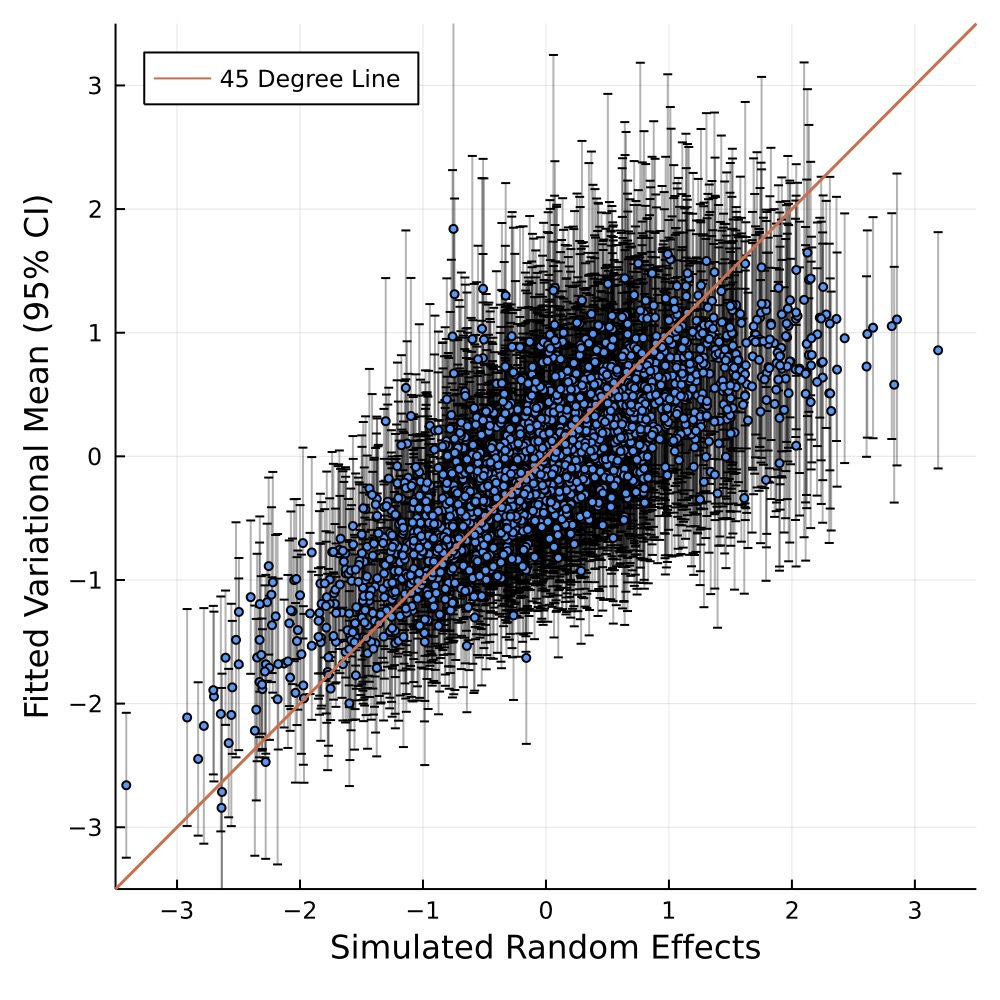}
    \end{subfigure}
    
    \begin{subfigure}{0.32\textwidth}
        \includegraphics[width=\textwidth]{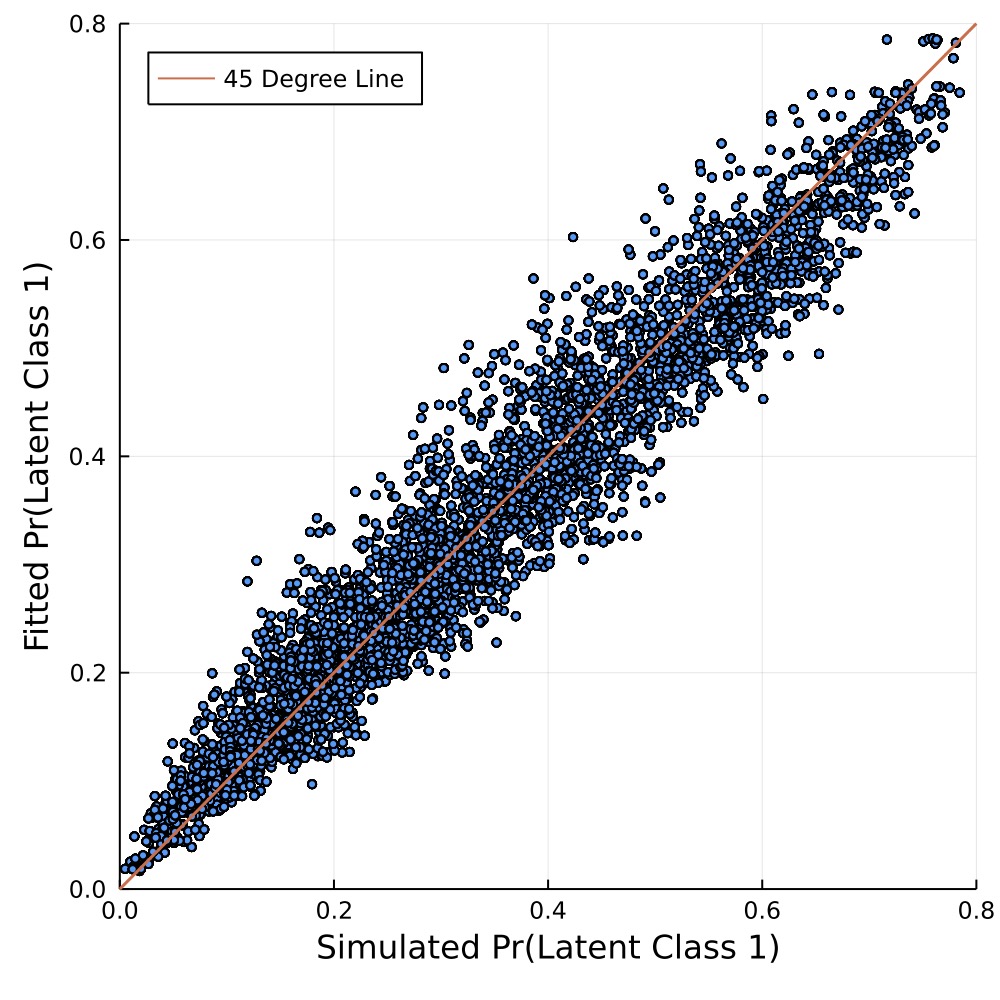}
    \end{subfigure}
    ~
    \begin{subfigure}{0.32\textwidth}
        \includegraphics[width=\textwidth]{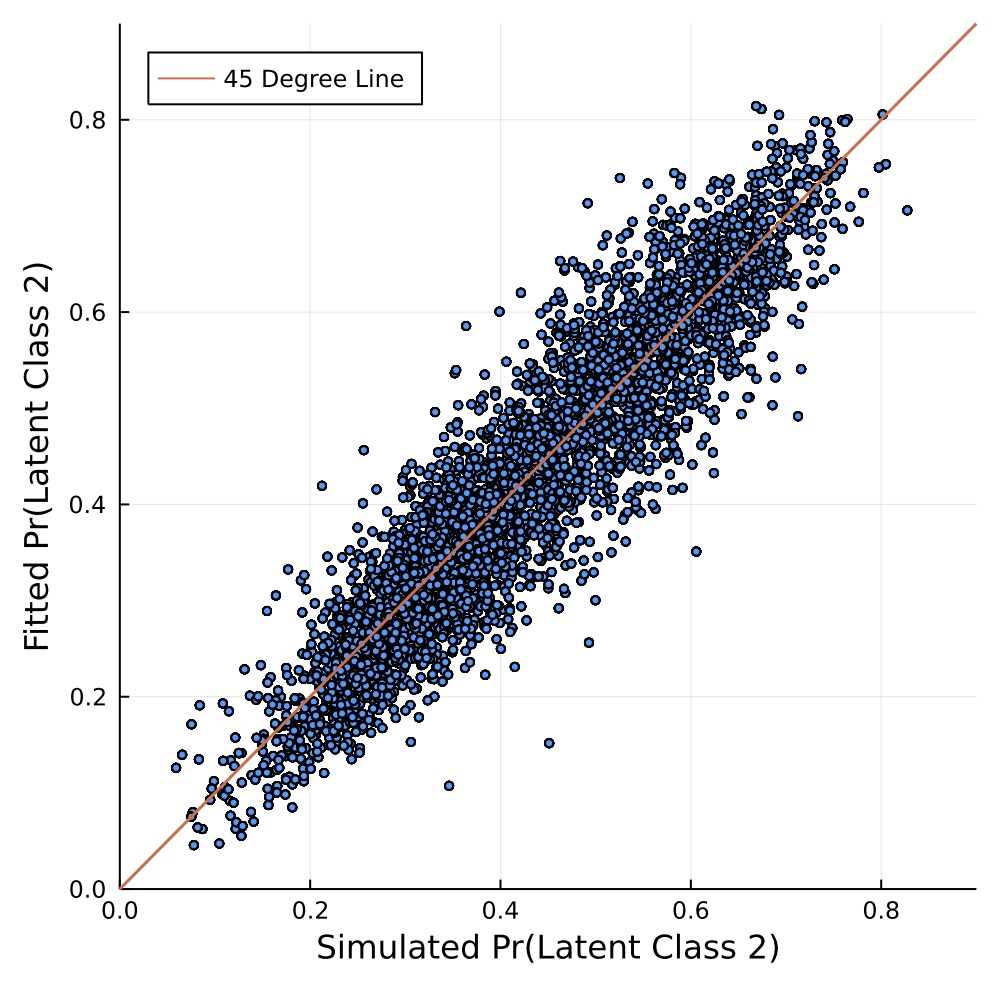}
    \end{subfigure}
    ~
    \begin{subfigure}{0.32\textwidth}
        \includegraphics[width=\textwidth]{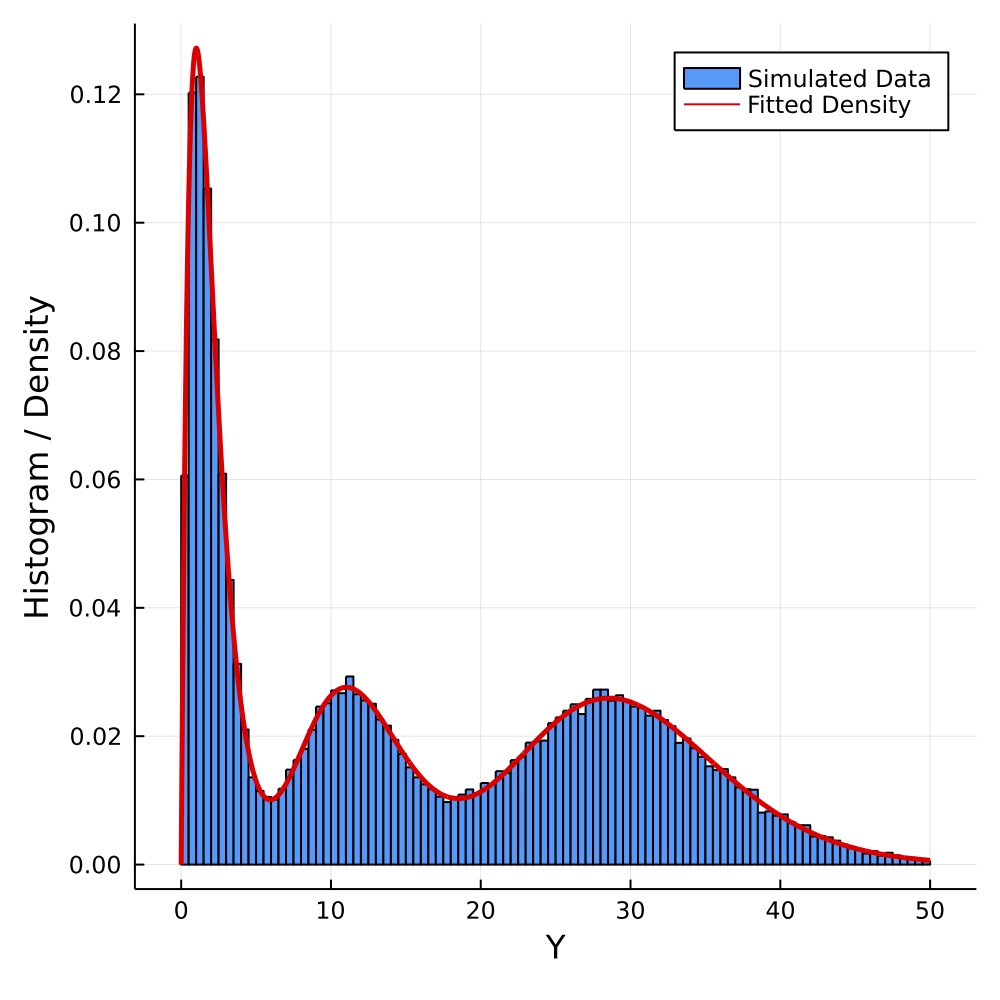}
    \end{subfigure}
\end{figure}

\pagebreak
}

In this section, we present two simulation studies in order to numerically illustrate the estimation algorithm described in \cref{sec:ParameterEstimation}. Our goal is to examine whether the proposed algorithm can correctly estimate the model parameters and make reasonable inference about the posterior distribution of random effects.

In both simulation studies, we consider a sample size of 50,000 observations where the covariate $\bm{x}_i$ consists of an intercept term $x_{i0} = 1$ and an indicator variable $x_{i1} \sim \textrm{Bernoulli}(0.5)$. The one-dimensional response $y_i$ is generated from a mixture of Gamma distributions, where the number of mixture components is two for case I and three for case II. As for the random effects, Simulation I contains one single level $\{w_{1}^{(s)}\}_{s=1, 2, \dots, 200}$ with $S_1=200$ levels randomly assigned to all observations. Simulation II has a more complex nested structure with two random effects such that $\{w_{1}^{(s)}\}_{s=1, 2, \dots, 200}$ has $S_1=200$ levels and $\{w_{2}^{(s)}\}_{s=1, 2, \dots, 2000}$ has $S_2=2000$ levels, both randomly assigned to all observations.

The true and fitted model parameters are summarized in \cref{table:simulation-1} and \cref{table:simulation-2}, while \cref{fig:simulation-1-RE} and \cref{fig:simulation-2-RE} visualize the simulated versus fitted random effects, latent class probabilities and the marginal distribution of response. Overall, we observe that the estimation algorithm is able to recover the true model parameters $(\bm{\alpha}, \bm{\beta}, \bm{\Psi})$ to a reasonable degree, which results in a close fit to the marginal distribution of the response variable, as indicated by the fitted density and the histogram of simulated data.

In addition to the response, we also investigate the simulated values versus the fitted posterior distribution of the random effects. We examine how well the approximated posterior credible intervals (CI) at different levels (90\%, 95\%, 97.5\% and 99\%) can recover the simulated true values of the random effects, which are summarized in the same set of tables and figures. In Simulation I, the random effects are well-recovered, and the plot of 95\%-CI shows a high level of alignment with the simulated true values of random effects. The results in Simulation II are noticeably worse but still acceptable, considering the added noises from two different types of random effects and much fewer observations per level. For example, each $w_{2}^{(s)}$ has 25 observations per factor on average, compared with 250 for $w_{1}^{(s)}$, which results in $w_{2}^{(s)}$ having much wider 95\%-CI and more cases where the posterior CI does not recover the true simulated values. Still, our algorithm is able to reasonably recover the latent class probabilities in both simulations, which contributes to the nice fit to the marginal distribution of the response.

For comparison, we have experimented the BLUP procedure outlined in \cite{yau2003finite} and \cite{ng2007extension} for similar MoE models with random effects. However, the BLUP procedure fails to recover the realizations of random effects, and we arrived at fitted models without any random effects (i.e.~all of $\bm{w}$ have degenerated to zero). Compared with the alternative of MCMC methods, our estimation algorithm is highly efficient in terms of computational cost. We have only used 50 iterations of ECM in both simulations to produce the results above, where $M=5$ samples of random effects are used in each ECM iteration for numerical evaluation such as \cref{eq:EZ-sampling-with-MC}. When implemented as a modification of the LRMoE.jl package, the computation time is 5 minutes for Case I and 15 minutes for Case II on a modern MacBook. We have experimented with standard MCMC algorithms as a benchmark, but our implementation did not converge within an acceptable time frame. A comparable study for MCMC methods in GLMM can be found in \cite{hadfield2010mcmc}, where the author analyzed a dataset with 828 observations and one single random effects with 106 levels. Their example converges with 60,000 total iterations, 10,000 iterations of burn-in and a thinning interval of 25. Considering our simulation studies are done on a much larger scale, we therefore reasonably expect our VI algorithm to be much more efficient than a comparable implementation of MCMC, in terms of the number of ECM iterations needed to converge. This computational advantage will be more significant in our real data analysis presented in the next section, where the number of levels of the random effect is more than a few thousands and our algorithm typically converges within two days after a few hundreds iterations of ECM.

\section{Real Data Analysis} \label{sec:RealDataAnalysis}

\afterpage{
\renewcommand{\arraystretch}{1.5}

\subfile{tables/real-data-overview-table}

\renewcommand{\arraystretch}{1}
}

In this section, we apply the Mixed LRMoE model to a real automobile insurance dataset for \emph{a posteriori} risk classification and ratemaking, and then compare its performance with a number of benchmark models. More specifically, we will investigate whether the Mixed LRMoE model can outperform benchmark models like GLM, GLMM and LRMoE without random effects in terms of goodness-of-fit. We will also investigate whether the Mixed LRMoE produces reasonable results for \emph{a posteriori} risk classification and ratemaking, that is, policyholders who made claims in the past should generally be considered riskier and should be assigned a higher \emph{a posteriori} premium.

The dataset contains the Bodily Injury (BI) claim history of 15,492 unique policyholders from policy years 2014 to 2019 (92,952 records in total) of a major North American automobile insurer. For illustration purposes, we have filtered for policyholders with exactly 6 years of history from 2014 to 2019. Practical issues, such as working with policyholders with a shorter history, or people with fractional policy year exposures, are trivial to address in the same modelling framework. Since we are only working with a one-dimensional response, it will be represented by $y_i$ in this section. The description of available covariates $\bm{x}_i$ and the summary statistics of the response $y_i$ are given in \cref{table:real-data-overview}. We observe the loss distribution has significant zero inflations and a heavy tail. We divide the entire dataset into training (2014--2018, or 12,394 unique policyholders with 61,968 records), validation (2014--2018, or 3,098 unique policyholders with 15,490 records) and testing (2019, for all 15,492 policyholders in training and validation) sets. Our goal is to fit various model candidates to the 5-year training period and then conduct \emph{a posteriori} risk classification and rakemaking for the 1-year testing period. The validation set contains a 20\% of the unique policyholders randomly selected from the 5-year training period, which is used for selecting the number of latent classes in the (Mixed) LRMoE models.

For illustration purposes, we will model the total amount of loss per year. As a benchmark, we will consider various combinations of GLM and GLMM against which we compare the proposed Mixed LRMoE model. For these benchmark models, we assume independence between claim frequency and severity. We use a probability mass $\delta_{i}(\bm{x}_i)$ at zero for no occurrence of claims and a continuous distribution $g_{i}(y_{i}|\bm{x}_i)$ for the total loss amount given there is at least one claim. Consequently, using $I_{\{y_{i}=0\}}$ and $I_{\{y_{i}>0\}}$ as indicators for the occurrence of claims, the distribution of total loss of policyholder $i$ is given by
\begin{equation}
    f(y_{i}|\bm{x}_i) = \delta_{i}(\bm{x}_i) \times I_{\{y_{i}=0\}} + (1-\delta_{i}(\bm{x}_i)) g_{id}(y_{i}|\bm{x}_i) \times I_{\{y_{i}>0\}}, \quad i=1, 2, \dots, n.
\end{equation}
where both $\delta_{i}(\bm{x}_i)$ and $g_{i}(y_{i}|\bm{x}_i)$ may be modelled by either GLM or GLMM. In the case of GLMM, we will add policyholder-level random effects with $15,492$ levels which corresponds to the number of unique policyholders in the training dataset.

For the models to investigate, we will consider (mixed) LRMoE with zero-inflated (ZI) lognormal expert functions. With the expert functions fixed, we only need to select the number of latent components for both LRMoE and Mixed LRMoE. We have selected a 4-component LRMoE and a 5-component Mixed LRMoE based on the Akaike Information Criterion (AIC) calculated on the validation dataset. For comparison with LRMoE without random effects, we also include a 4-component Mixed LRMoE in the following discussion.

\subsection{Goodness-of-Fit} \label{sec:goodness-of-fit}

\afterpage{
\renewcommand{\arraystretch}{1.5}

\subfile{tables/real-data-model-benchmark}

\vfill

\subfile{tables/real-data-model-LRMoEs}

% \vfill
% \subfile{tables/real-data-zero-prob}
\renewcommand{\arraystretch}{1}
\vfill
\begin{figure}[!h]
    \centering
    \caption{Histogram and fitted density of positive claim distribution. Left / Right: Training / Testing set.}
    \label{fig:real-data-fitted-vs-histogram}
    \begin{subfigure}{0.32\textwidth}
        \includegraphics[width=\textwidth]{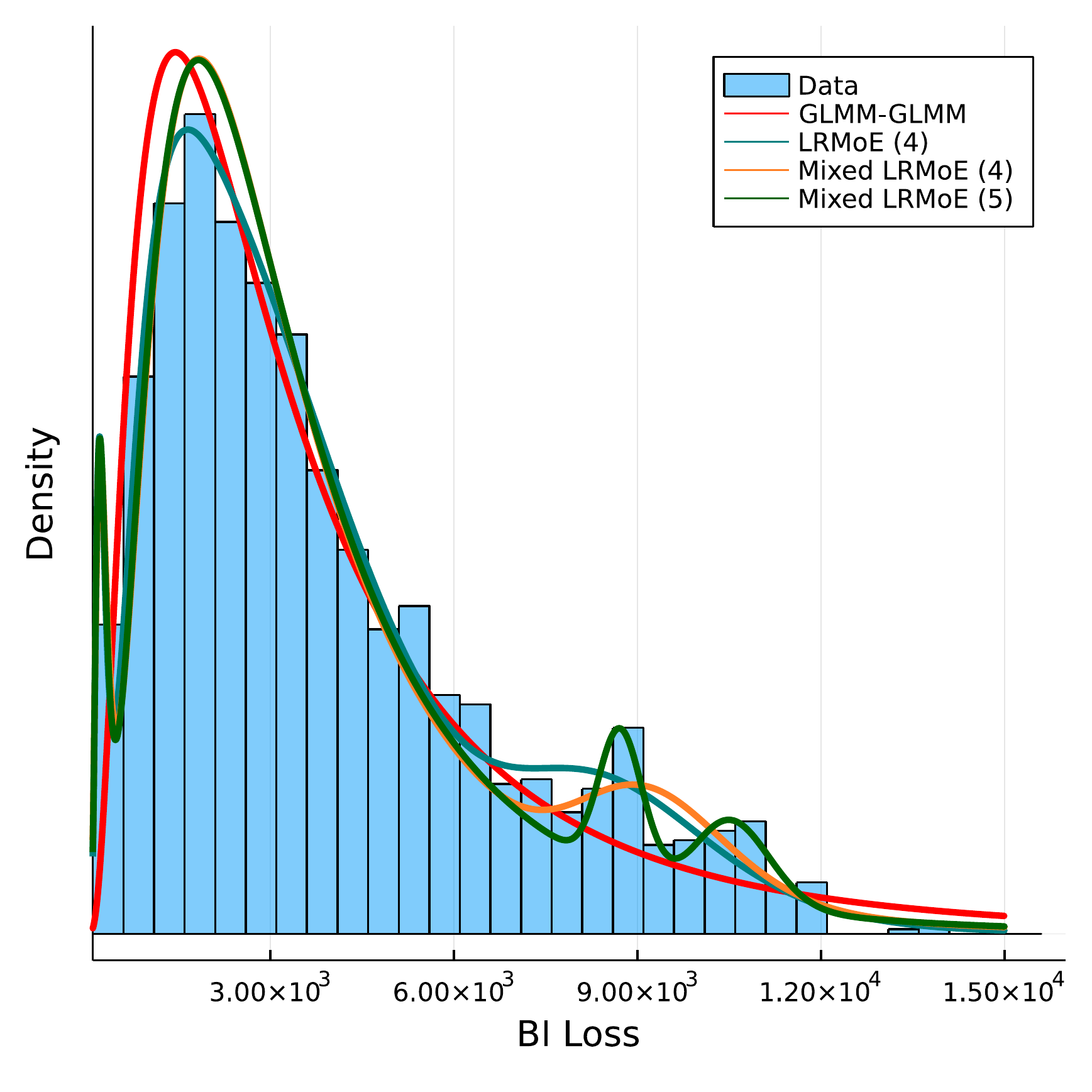}
    \end{subfigure}
    ~
    \begin{subfigure}{0.32\textwidth}
        \includegraphics[width=\textwidth]{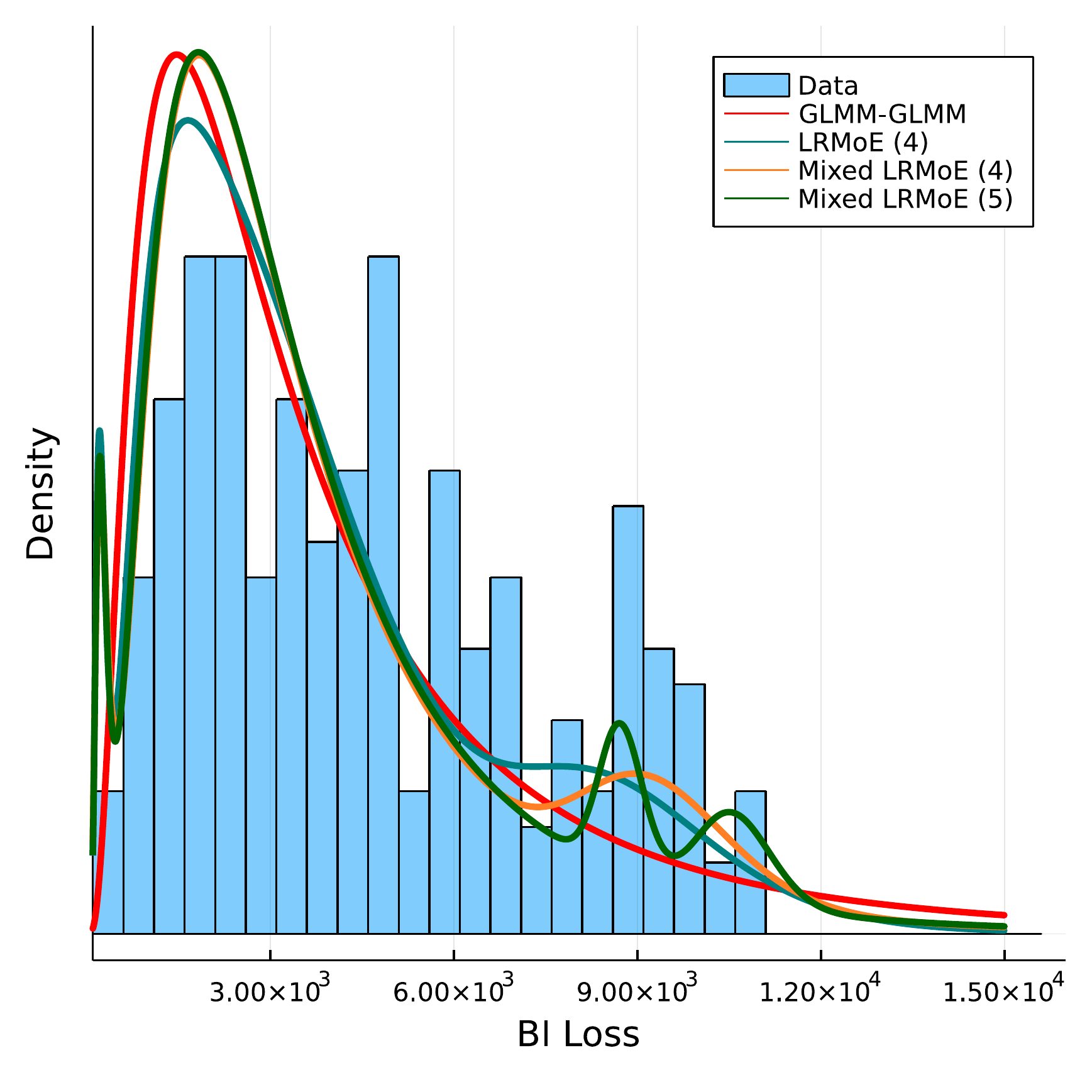}
    \end{subfigure}
\end{figure}
\pagebreak
}

The fitted loglikelihood values of all benchmark models are summarized in \cref{table:real-data-model-benchmark}. As expected, the GLMM-GLMM model produces the highest loglikelihood since the policyholder-level random effects are used twice. The combinations of GLMM-GLM and GLM-GLMM offer slightly worse fit to data, followed by the GLM-GLM model without any random effects. \cref{table:real-data-model-LRMoEs} summarizes the loglikelihood of the 4-component LRMoE and two Mixed LRMoE models. We see that both mixed LRMoE models offer much better fit to data in terms of loglikelihood on training and testing datasets, and both outperform the LRMoE model without random effects. This demonstrates the flexibility of mixed LRMoE as well as the advantage of incorporating policyholder-level random effects for more accurate modelling of the loss distribution. As for penalization on model complexity, we also include the number of parameters for all model candidates in the tables. It is clear that the Mixed LRMoE models outperforms all benchmark models in terms of AIC on the training set, while they have marginally worse AIC values on the testing set. However, as will be evident in the next subsection, this added model complexity greatly improves \emph{a posteriori} risk classification and ratemaking, which is the ultimate goal in this context.

Besides loglikelihood values, we also look at how each model candidate fits the probability of claim and the distribution of positive losses. For the probability of claim, all model candidates offer very similar fitting performance. On the training period, all models are able to fit the observed claim probability 0.978996 to the fourth decimal place. However, on the testing period where the observed claim probability is 0.988793, all models candidates have produced a slightly lower prediction, ranging from 0.979844 to 0.980005 (or $-0.91\%$ to $-0.89\%$ of relative error). Meanwhile, the (Mixed) LRMoE models have provided a better fit to the distribution of positive losses, as indicated by \cref{fig:real-data-fitted-vs-histogram} which compares the fitted densities against the empirical distribution. Most notably, the (Mixed) LRMoE models have successfully captured the multimodality in the tail, while GLM and GLMM only fit a unimodal density to the entire distribution of positive losses.

For both the claim probability and the distribution of positive losses, we have observed a potential data drift for testing period. In particular, the claim probability increases in 2019 compared with previous years, while the distribution of positive losses also appears to have changed a little, but the latter is only based on roughly 170 losses observed in the testing period.

\subsection{Risk Classification and Ratemaking}

\afterpage{
\renewcommand{\arraystretch}{1.5}

\subfile{tables/real-data-prob-changes}

\renewcommand{\arraystretch}{1}
}

\afterpage{
\begin{figure}[!ht]
    \centering
    \caption{Histogram of predicted posterior premium based on different models. Top row, from left to right: GLM-GLM, GLM-GLMM, GLMM-GLM, GLMM-GLMM. Bottom row, from left to right: LRMoE (4), Mixed LRMoE (4), Mixed LRMoE (5).}
    \label{fig:real-data-contrast-premium}
    
    \begin{subfigure}{0.23\textwidth}
        \includegraphics[width=\textwidth]{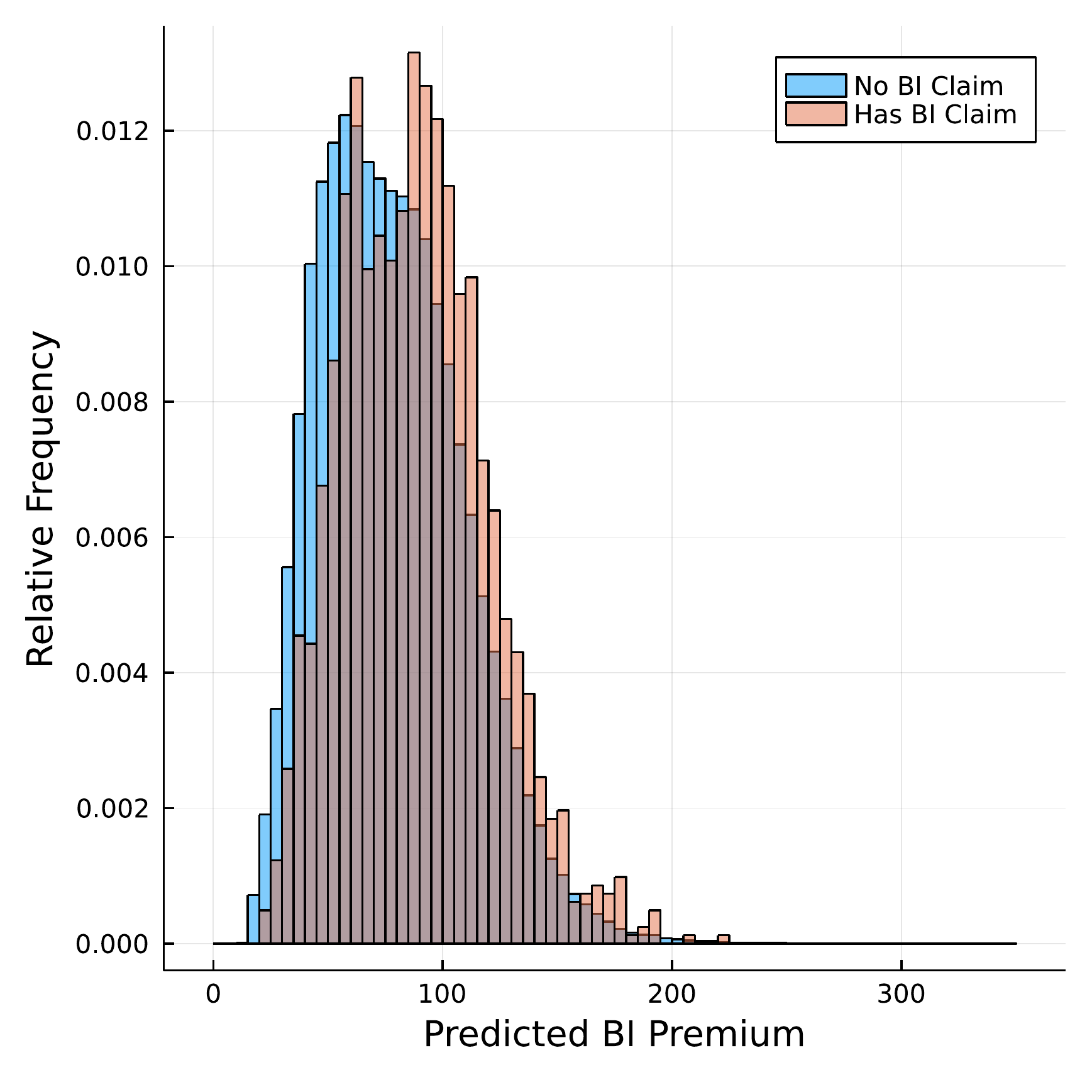}
    \end{subfigure}
    ~
    \begin{subfigure}{0.23\textwidth}
        \includegraphics[width=\textwidth]{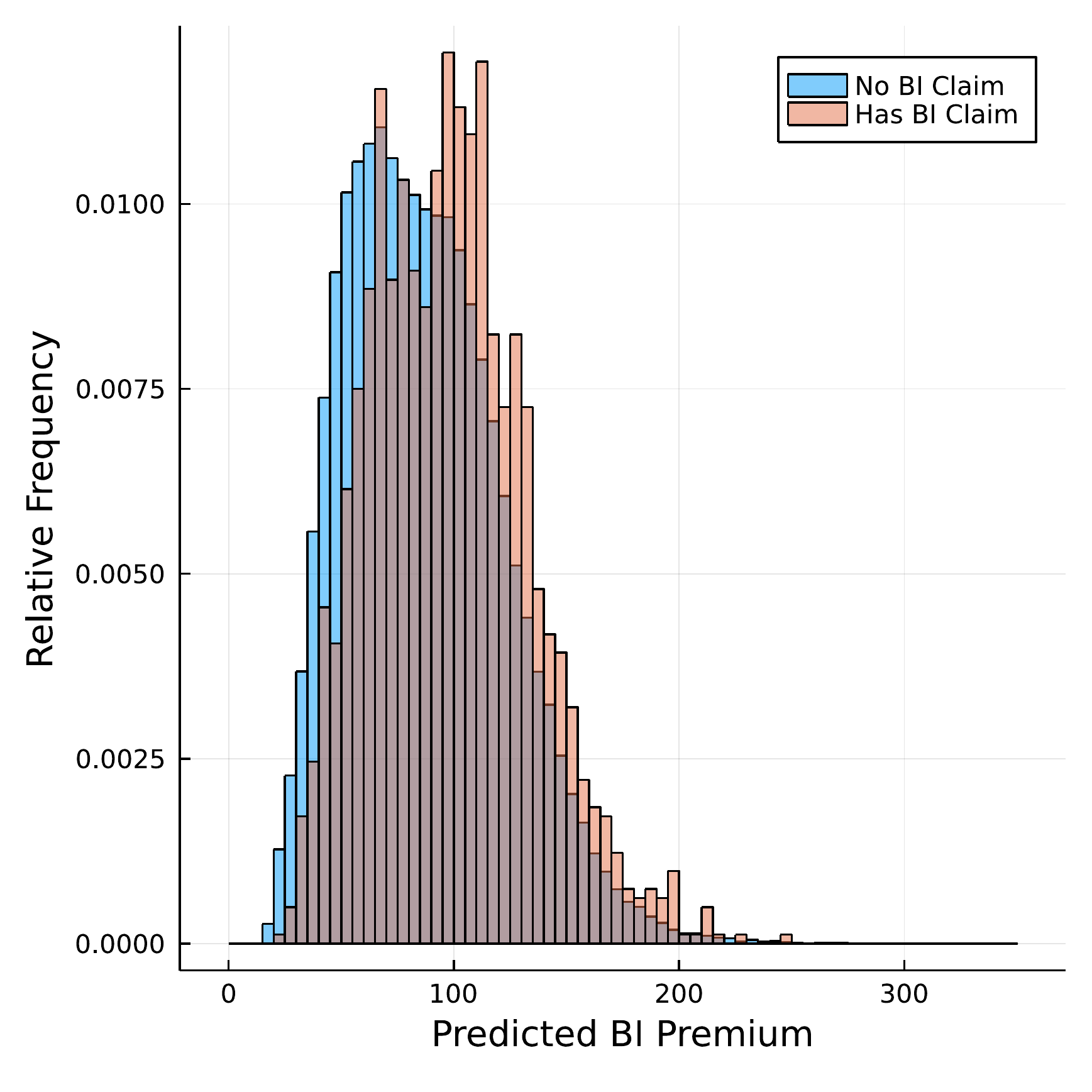}
    \end{subfigure}
    ~
    \begin{subfigure}{0.23\textwidth}
        \includegraphics[width=\textwidth]{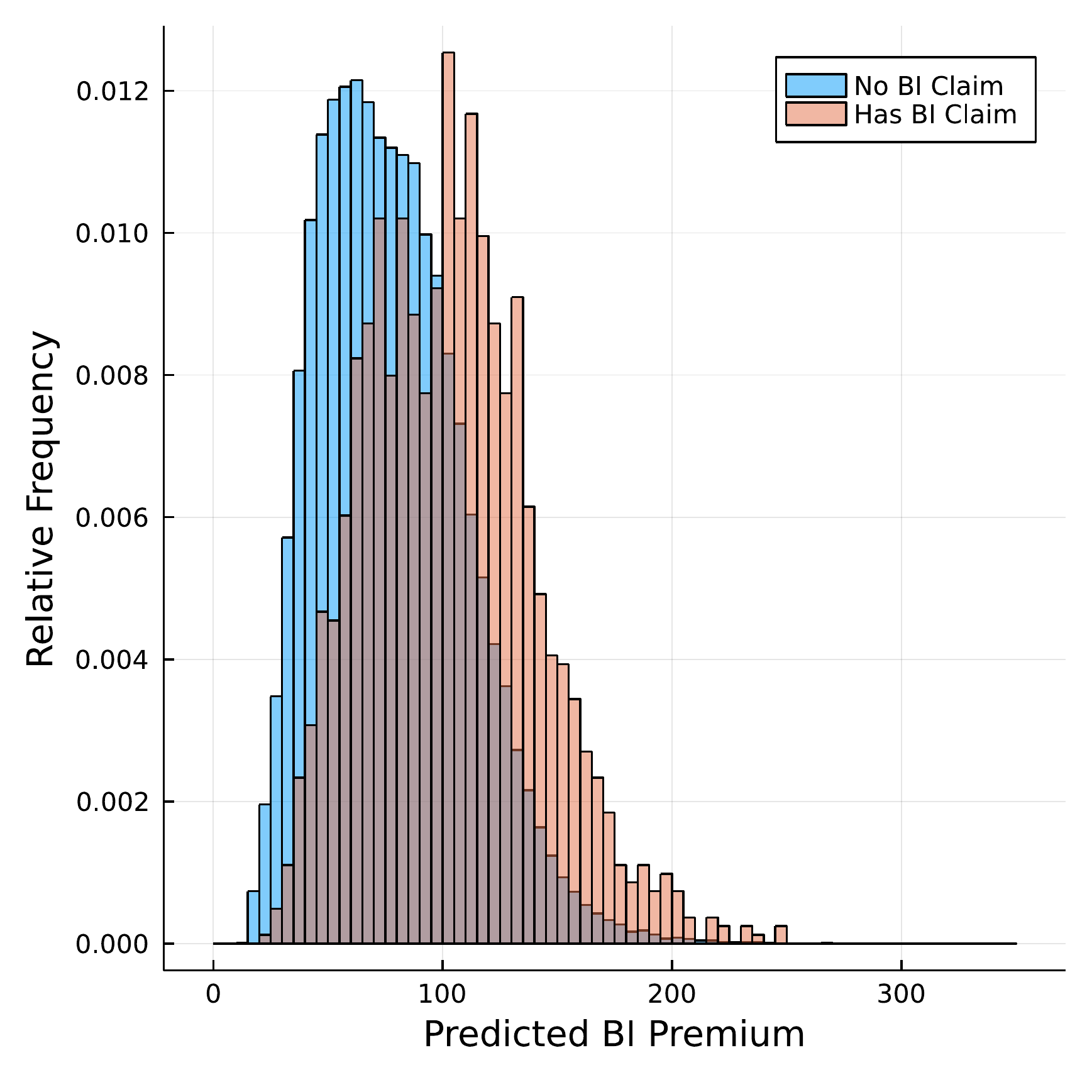}
    \end{subfigure}
    ~
    \begin{subfigure}{0.23\textwidth}
        \includegraphics[width=\textwidth]{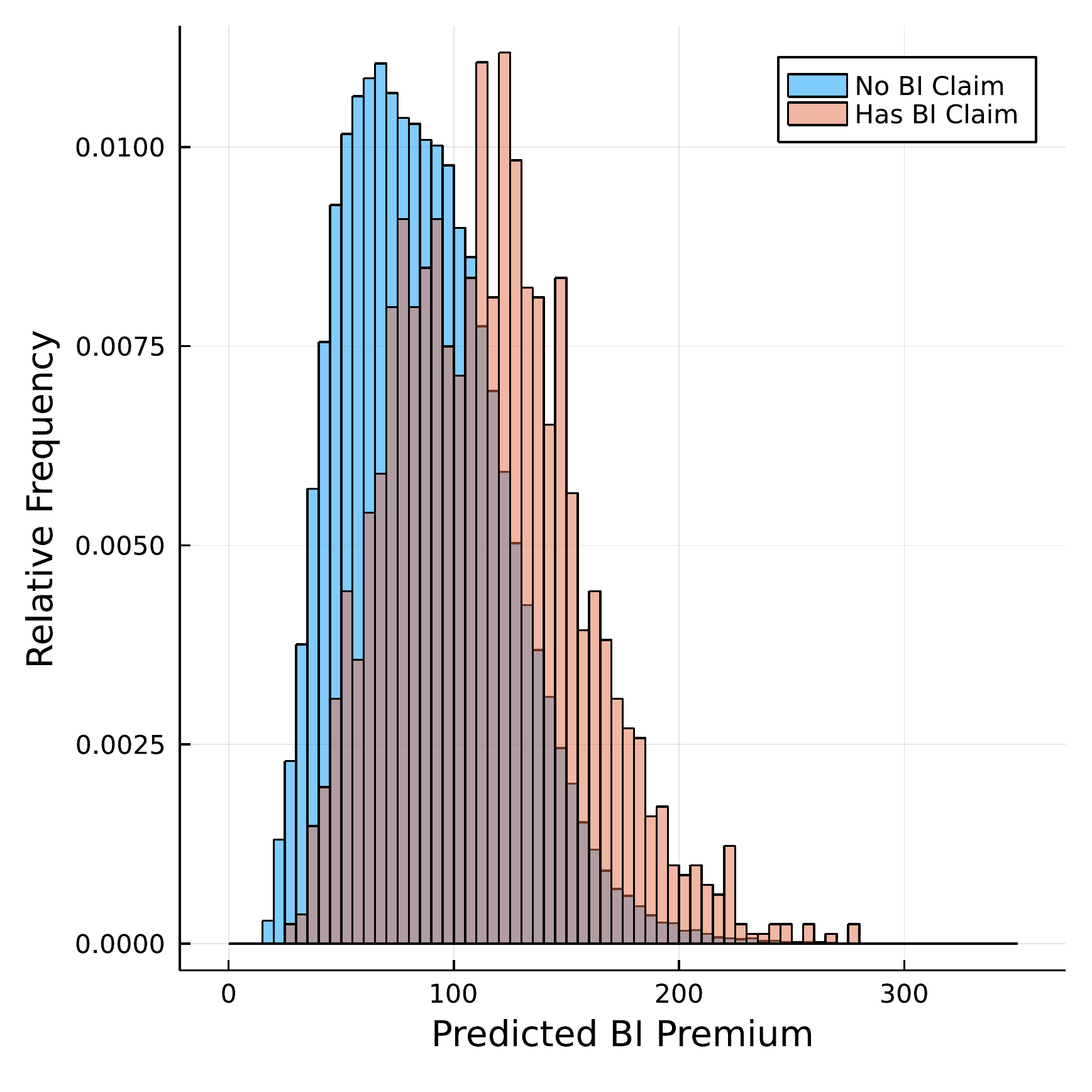}
    \end{subfigure}

    \begin{subfigure}{0.23\textwidth}
        \includegraphics[width=\textwidth]{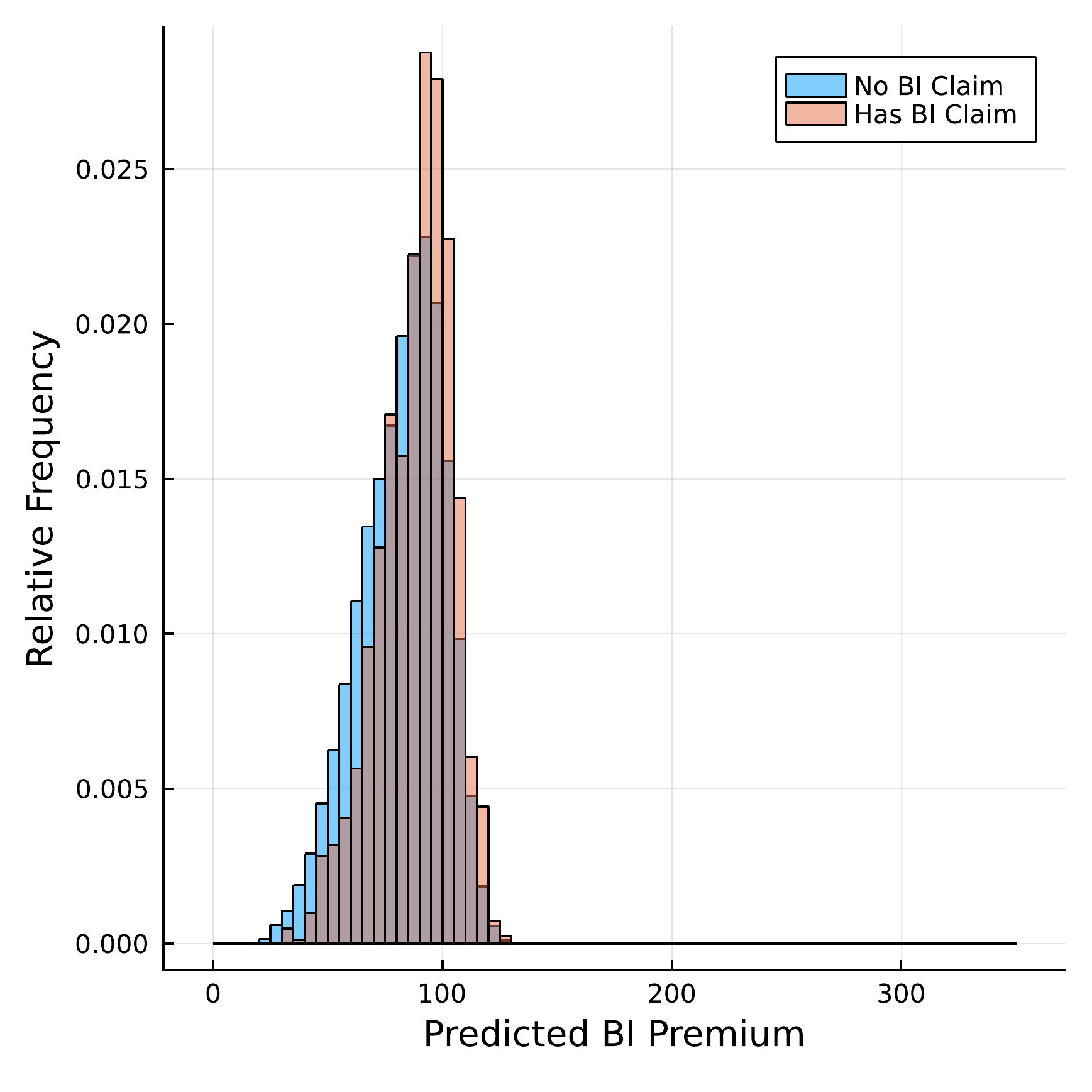}
    \end{subfigure}
    ~
    \begin{subfigure}{0.23\textwidth}
        \includegraphics[width=\textwidth]{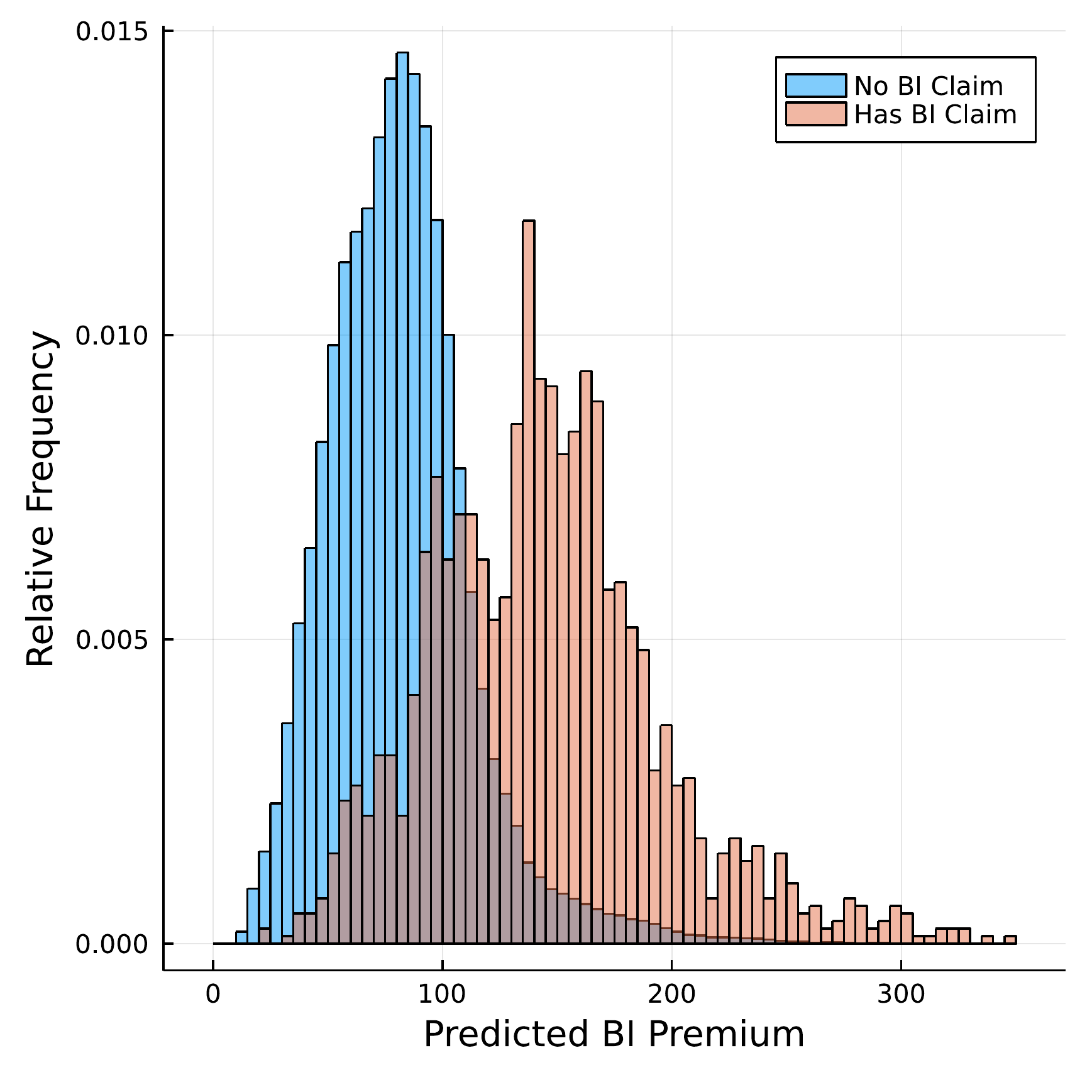}
    \end{subfigure}
    ~
    \begin{subfigure}{0.23\textwidth}
        \includegraphics[width=\textwidth]{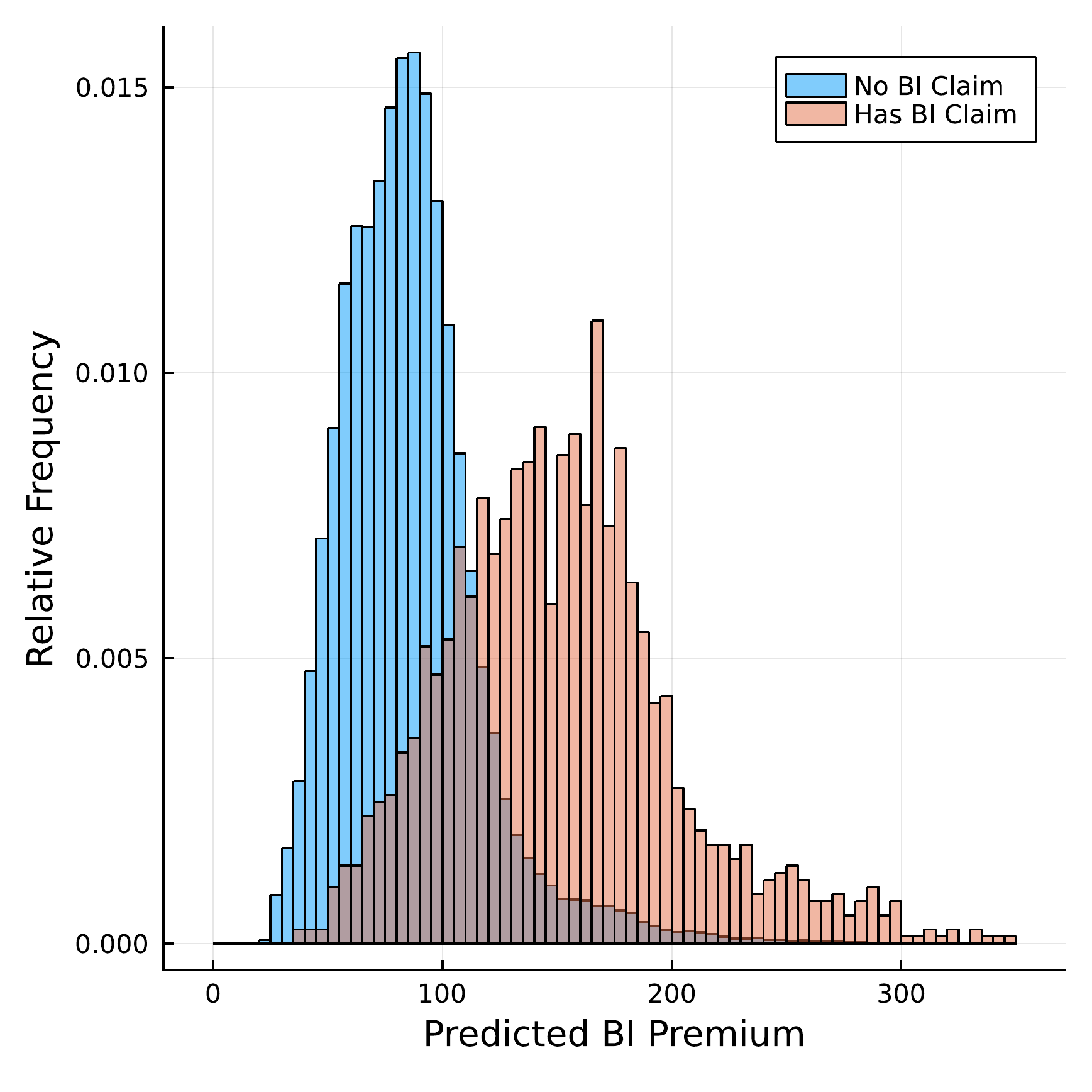}
    \end{subfigure}
\end{figure}

\renewcommand{\arraystretch}{1.5}

\subfile{tables/real-data-premium-table}

\renewcommand{\arraystretch}{1}

\pagebreak
}

For insurance pricing purposes, it is crucial that policyholders' claim history is adequately incorporated in the calculation of premium at policy renewal. In short, higher risks, as reflected by the occurrence of claim and/or higher claim amounts, should lead to a higher \emph{a posteriori} premium. In this subsection,
we compare the model performance in terms of \emph{a posteriori} risk classification and ratemaking.

For risk classification, the latent classes in (mixed) LRMoE models can be naturally interpreted as different clusters of policyholders based on their risk profile. To compare how risk classification is affected by claim history, we categorize all policyholders into two groups: those with at least one claim and those without any claim during 2014--2018, and summarize their latent class probabilities in \cref{table:real-data-prob-changes}. Most notably, with the addition of random effects, the Mixed LRMoE models are able to strongly distinguish risky policyholders who have at least one claim in the past, by assigning almost double the probability to the riskiest latent class. Meanwhile, the LRMoE model without random effects only suggests a slight increase in the risky class probability based solely on covariate information, given the independence assumption for observations across different policy years.

Different decisions in \emph{a posteriori} risk classification will also lead to differences in ratemaking. For \emph{a posteriori} ratemaking, we calculate the premium for policy renewals in year 2019 based on the posterior distribution given the claim history in 2014--2018. For illustration purposes, we only consider the pure premium which is equal to the probability of claim multiplied to the expected positive mean loss amount.

On a higher level, we investigate all policyholder based on the same grouping (with and without claims in 2014--2018). The distributions of the predicted posterior premium are shown in \cref{fig:real-data-contrast-premium} for all model candidates. For models without random effects, i.e.~GLM-GLM and LRMoE, the predicted distributions of posterior premium for the two groups appear to be highly overlapping, which indicates that fixed effects alone cannot distinguish policyholders based on claim history. For benchmark models with random effects, namely GLM-GLMM, GLMM-GLM and GLMM-GLMM, there appears to be some difference between the two groups, whereby some policyholders with claim history will have a higher predicted premium. Most notably, the two Mixed LRMoE models show much larger differences between the distributions of predicted premium, which better captures the riskiness of policyholders reflected by their claim history.

On a more detailed level, \cref{table:real-data-comparison-premium-table} summarizes the predicted posterior premium, based on the two groups above in addition to the relative size of incurred total losses. We observe that both Mixed LRMoE models, as well as benchmark GLMM-GLM and GLMM-GLMM, heavily penalizes policyholders who have at least one claim, as shown by the additional premium loadings. For people with claims, only the Mixed LRMoE with five components has further provided a correct ordering of the predicted posterior premium, that is, people with larger incurred claims typically have higher premium at policy renewal, which indicates better performance in \emph{a posteriori} risk classification and ratemaking. This is because only the 5-component Mixed LRMoE has adequately captured the multimodality in the tail of the positive loss distribution, as indicated by \cref{fig:real-data-fitted-vs-histogram}.

For both \emph{a posteriori} risk classification and ratemaking discussed above, we have primarily focused on differentiating policyholders based on the occurrence of claims and the claim sizes when applicable, whereby the Mixed LRMoE models are shown to have effectively incorporated such information. However, we can still observe the effects of \emph{a priori} information, i.e.~policyholder covariates, when determining the \emph{a posteriori} premium. Most notably, in \cref{fig:real-data-contrast-premium}, there is a good level of overlap between the histograms of the predicted premium for people with and without claim history, even for all model candidates with random effects. For example, certain policyholders with claim history (lower end of the orange histogram) would still be charged a lower premium than some policyholders without claim history (upper end of the blue histogram), which should be attributed to covariates such as the inherent risk level of certain age groups or the collision rating of a particular group of vehicles.

\subsection{Gini Index}

\afterpage{
\renewcommand{\arraystretch}{1.5}
\begin{figure}[!h]
    \centering
    \caption{Comparison of the Ordered Lorenz Curves generated from various model candidates. Left / Right: Training / Testing set.}
    \label{fig:real-data-contrast-gini-index}
    
    \begin{subfigure}{0.32\textwidth}
        \includegraphics[width=\textwidth]{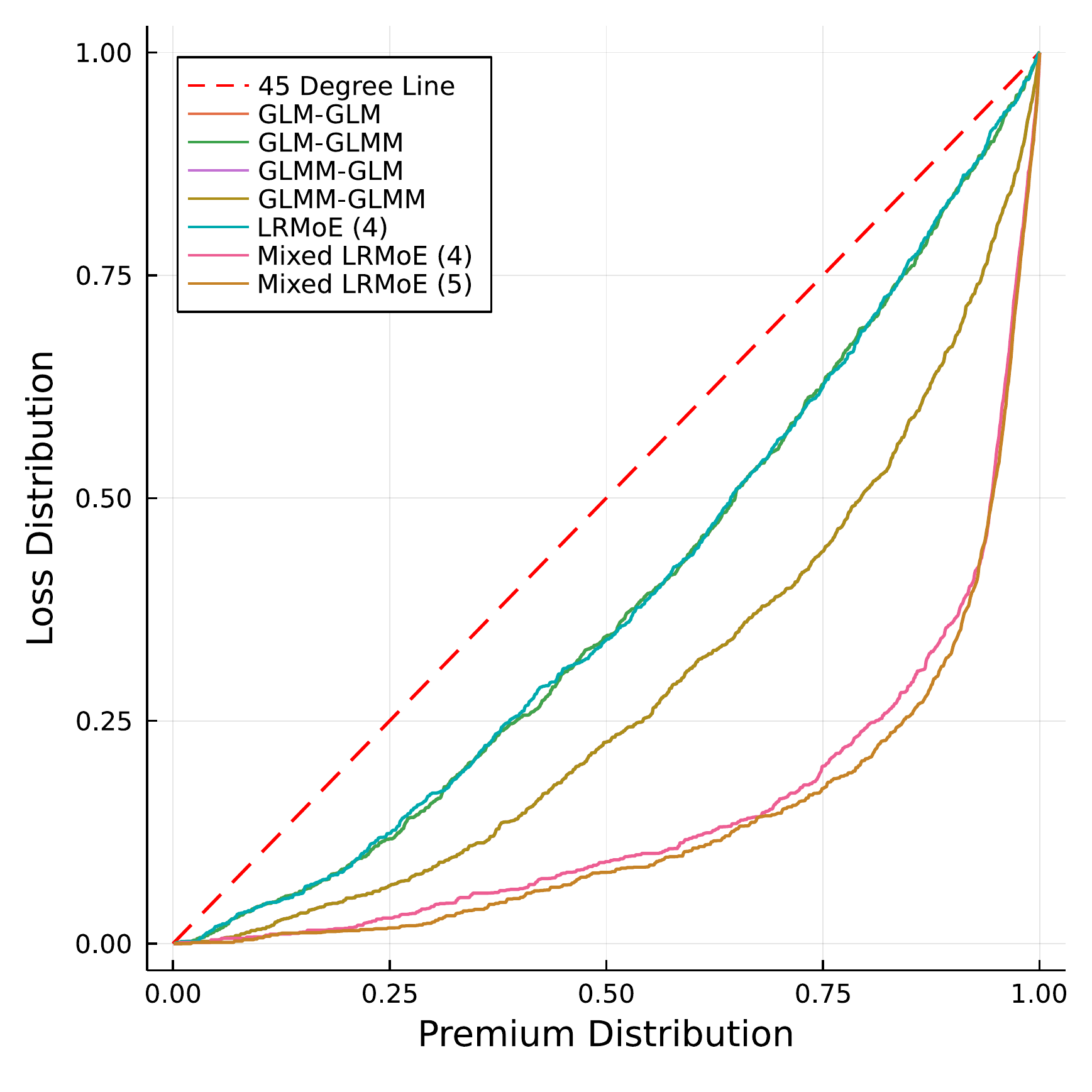}
    \end{subfigure}
    ~
    \begin{subfigure}{0.32\textwidth}
        \includegraphics[width=\textwidth]{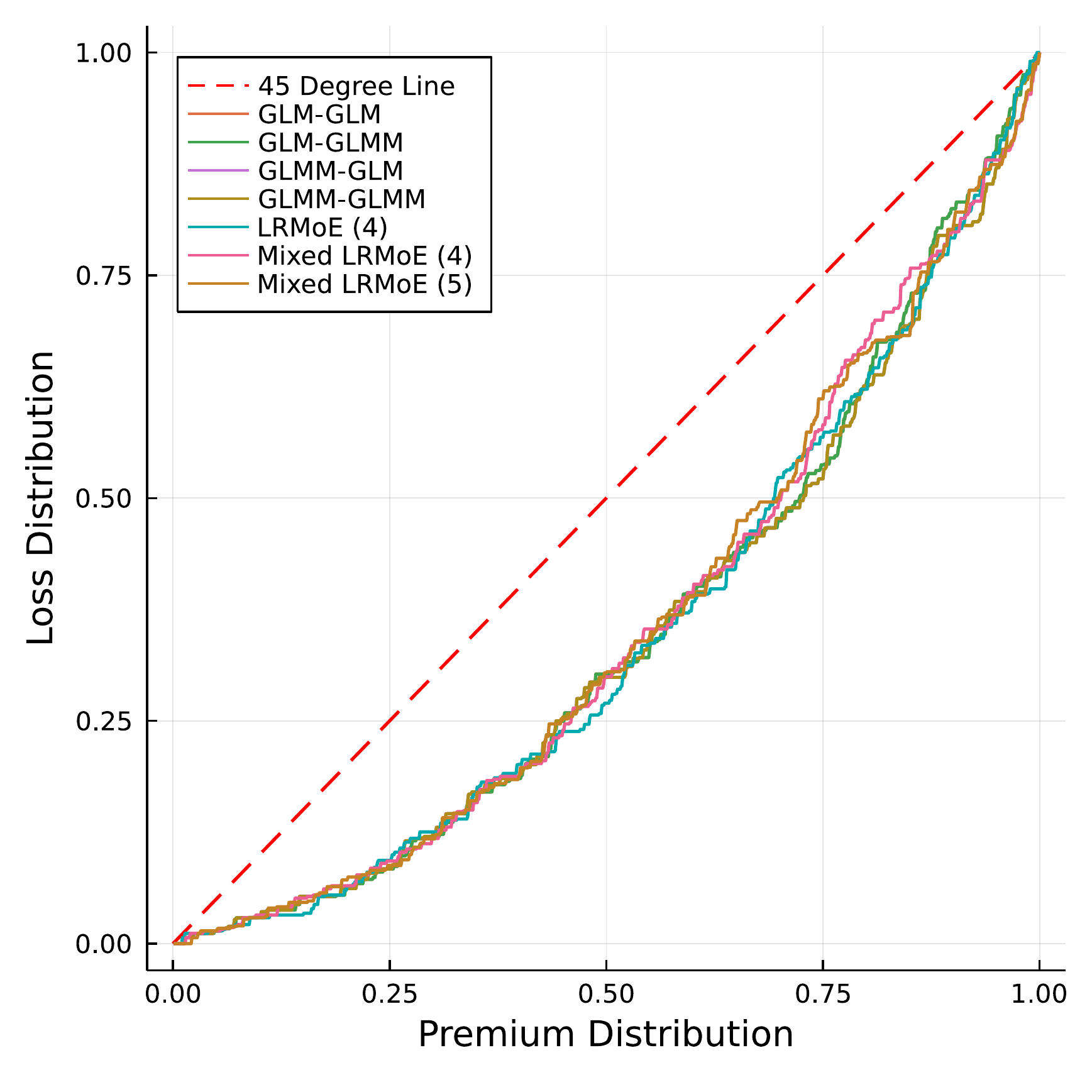}
    \end{subfigure}
\end{figure}

\subfile{tables/real-data-Gini-comparison}

\renewcommand{\arraystretch}{1}
}

Finally, we examine the model performance using the Gini Index as a measurement of adequacy for insurance risk scoring (see e.g.~\cite{frees2011summarizing}).
We first plot the Ordered Lorenz Curve in \cref{fig:real-data-contrast-gini-index} for both the training and testing sets, where the $x$-axis represents the cumulative percentage of premium and $y$-axis represents the cumulative percentage of the incurred losses during the training or testing period. The corresponding Gini index values for all model candidates, calculated as twice the area between the Ordered Lorenz Curve and the Line of Equality (45-degree line), as well as their estimated standard error, are summarized in \cref{table:real-data-comparison-Gini}.

On the training set, we see the two mixed LRMoE models have produced Ordered Lorenz Curves farthest from the Line of Equality as well as the largest Gini Index values, which indicates a high degree of differentiation between low- and high-risk policyholders based on their claim history. We also note the second best models in terms of Gini index are GLMM-GLM and GLMM-GLMM, which means the probability of claim may potentially be a more important determinant of policyholders' risk profile compared with claim severity.

However, on the testing set, all model candidates perform quite similarly, and the two mixed LRMoE models do not outperform the classical models. In fact, the estimated standard errors of the Gini Index suggest we cannot conclude whether the performances of all model candidates are significantly different from each other. This may have been caused by the small number of incurred claims, as observed in \cref{sec:goodness-of-fit}, but a more important factor might be the potential data drift in year 2019 with a slightly lower claim probability and changed distribution of positive losses. However, such unprecedented data drift is outside the scope of what statistical and predictive models can address based on historical data only.

\section{Conclusion} \label{sec:Conclusion}

In this paper, we have proposed to incorporate policyholder-level random effects in a flexible regression framework, called the Mixed LRMoE, which is then applied to the Bonus-Malus problem. Although the addition of random effects has resulted in an intractable marginal likelihood function of the model, we have developed a stochastic variational ECM algorithm for efficient estimation of model parameters and inference of the posterior of random effects, which are crucial for updating policyholders' risk profile based on their claim history. Our numerical simulation and real data analysis have demonstrated the potentials of Mixed LRMoE as a powerful tool for more accurate insurance loss modelling and better \emph{a posteriori} insurance risk classification and ratemaking. While our current work has already shown promising results, one may consider the following extensions and directions for future work.
\begin{itemize}
    \item In the current formulation of Mixed LRMoE, all past policy years are equally weighted by sharing the same realization of random effects. A more realistic and general approach is to apply a weighting scheme whereby recent claims are more influential in determining the posterior premium.
    \item We have taken the approach of modelling the total incurred loss as a mixture of zero-inflated distributions, whereby the dependence between claim frequency and severity are not explicitly specified. An interesting extension is to incorporate such dependence in the (Mixed) LRMoE modelling framework.
    \item While our estimation algorithm enjoys numerical efficiency and has been shown to yield reasonable results both in simulation and real data analysis, it could be worthwhile to investigate the theoretical properties, such as approximation errors and rate of convergence, of VI methods in the class of MoE models as well as the Mixed LRMoE.
\end{itemize}

\appendix

\section{Detailed Implementation of the Estimation Algorithm} 

\subsection{Stochastic Version of IRLS} \label{appendix:detailed-estimation-IRLS}

We continue with $Q_1^{(t+1)}(\bm{\alpha}, \bm{\beta}; \bm{X}, \bm{\Theta}^{(t)})$ in \cref{eq:Q1-objective-function-alpha-beta} which is to be maximized in $(\bm{\alpha}, \bm{\beta})$. For the IRLS algorithm, we implement the following steps that conditionally maximize the objective function to seek a local optimum.
\begin{enumerate}[label=(\arabic*)]
    \item Optimize $Q_1^{(t+1)}$ with respect to $\bm{\alpha}_1$, given the previous values of $\bm{\beta}_1^{(t)}$ and $(\bm{\alpha}_j^{(t)}, \bm{\beta}_j^{(t)})$ for $j=2, 3, \dots, g$.
    \item Optimize $Q_1^{(t+1)}$ with respect to $\bm{\beta}_1$, given the updated values of $\bm{\alpha}_1^{(t+1)}$ and the previous values $(\bm{\alpha}_j^{(t)}, \bm{\beta}_j^{(t)})$ for $j=2, 3, \dots, g$.
    \item Repeat (1) and (2) for $(\bm{\alpha}_2, \bm{\beta}_2)$, given the updated values of $(\bm{\alpha}_1^{(t+1)}, \bm{\beta}_1^{(t+1)})$ and the previous values $(\bm{\alpha}_j^{(t)}, \bm{\beta}_j^{(t)})$ for $j=3, 4, \dots, g$.
    \item Repeat (3) for all $j=1, 2, \dots, g-1$ until convergence.
\end{enumerate}

For step (1), given the realization of random effects $\bm{w}$, the optimal $\bm{\alpha}_j$ is updated by the following equation until convergence.
\begin{equation} \label{eq:IRLS-update-alpha}
    \bm{\alpha}_j \leftarrow \bm{\alpha}_j - \left[ \frac{\partial^2 Q_1}{\partial \bm{\alpha}_j \partial \bm{\alpha}_j^T} \right]^{-1} \frac{\partial Q_1}{\partial \bm{\alpha}_j}
\end{equation}
where
\begin{equation} \label{eq:dQ1dalpha}
\begin{aligned}
    \frac{\partial Q_1}{\partial \bm{\alpha}_j} &= \sum_{i=1}^{n} z_{ij}^{(t)} \left[ 1 - \frac{\exp( \bm{\alpha}_j^T \bm{x}_i + \bm{\beta}_j^T\bm{w}_i )}{ \sum_{j'=1}^{g} \exp( \bm{\alpha}_{j'}^T \bm{x}_i + \bm{\beta}_{j'}^T\bm{w}_i ) } \right] \bm{x}_i 
\end{aligned}
\end{equation}
and
\begin{equation} \label{eq:dQ12dalpha2}
    \frac{\partial^2 Q_1}{\partial \bm{\alpha}_j \partial \bm{\alpha}_j^T} = \sum_{i=1}^{n}  \frac{\exp( \bm{\alpha}_j^T \bm{x}_i + \bm{\beta}_j^T\bm{w}_i ) \left[ \exp( \bm{\alpha}_j^T \bm{x}_i + \bm{\beta}_j^T\bm{w}_i ) - \sum_{j'=1}^{g} \exp( \bm{\alpha}_{j'}^T \bm{x}_i + \bm{\beta}_{j'}^T\bm{w}_i ) \right] }{ \left[ \sum_{j'=1}^{g} \exp( \bm{\alpha}_{j'}^T \bm{x}_i + \bm{\beta}_{j'}^T\bm{w}_i ) \right]^2 }  \bm{x}_i \bm{x}_i^T.
\end{equation}

To marginalize over the random effects, \cref{eq:dQ1dalpha} and \cref{eq:dQ12dalpha2} are replaced with their Monte Carlo versions by sampling $\bm{w}$ from the variational distribution $\bm{q}(\cdot ;\bm{\Theta}^{(t)})$.

Step (2) is similarly carried out by noting the resemblance between $\bm{\alpha}_j^T \bm{x}_i$ and $\bm{\beta}_j^T\bm{w}_i$. Essentially, $\bm{\beta}_j$ is updated with the same method as \cref{eq:IRLS-update-alpha}, where the gradient and hessian matrix with respect to $\bm{\beta}_j$ are obtained by appropriately replacing $\bm{x}_i$ with $\bm{w}_i$ in \cref{eq:dQ1dalpha} and \cref{eq:dQ12dalpha2}. Details are omitted.

\subsection{Optimization over Variational Parameters} \label{appendix:detailed-estimation-VI-params}

Given the updated model parameters $(\bm{\alpha}^{(t+1)}, \bm{\beta}^{(t+1)}, \bm{\Psi}^{(t+1)})$, we aim to use gradient descent to maximize the following objective function with respect to the variational parameters $\bm{\Theta} = \{(\bm{\mu}_l, \bm{\Sigma}_l)\}_{l=1, 2, \dots, L}$.
\begin{equation} \label{eq:Q-in-VI-after-model-params}
\begin{aligned}
   Q^{(t+1)}(\bm{\Theta}; \bm{X}, \bm{Y}, \bm{\alpha}^{(t+1)}, \bm{\beta}^{(t+1)}, \bm{\Psi}^{(t+1)}) 
    &= \E_{\bm{w} \sim \bm{q}(\cdot;\bm{\Theta})} \left[ \tilde{\ell}^{c}(\bm{\alpha}^{(t+1)}, \bm{\beta}^{(t+1)}, \bm{\Psi}^{(t+1)}; \bm{X}, \bm{Y}, \bm{w}) \right] \\ &- \textrm{KL}\left[ \bm{q}(\bm{w} ; \bm{\Theta}) || \bm{\phi}(\bm{w}) \right]
\end{aligned}
\end{equation}
where
\begin{equation}
    \tilde{\ell}^{c}(\bm{\alpha}^{(t+1)}, \bm{\beta}^{(t+1)}, \bm{\Psi}^{(t+1)}; \bm{X}, \bm{Y}, \bm{w}) = \sum_{i=1}^{n} \sum_{j=1}^{g} z_{ij}^{(t)} \log \left[ \pi_{j}(\bm{x}_i, \bm{w}_i; \bm{\alpha}^{(t+1)}, \bm{\beta}^{(t+1)}) f_{j}(\bm{y}_i; \bm{\psi}_j^{(t+1)}) \right].
\end{equation}

Given the normality assumption of both $\bm{q}( \cdot ; \bm{\Theta})$ and $\bm{\phi}(\cdot)$, the KL divergence term is simplified as
\begin{equation}
     \textrm{KL}\left[ \bm{q}(\bm{w} ; \bm{\Theta}) || \bm{\phi}(\bm{w}) \right] = -\frac{1}{2} \sum_{l=1}^{L}\left[ - \log |\bm{\Sigma}_l| + \textrm{tr}(\bm{\Sigma}_l) + \bm{\mu}_l^T \bm{\mu}_l \right] + \textrm{const.}
\end{equation}
which yields closed-form gradient and hessian with respect to both $\bm{\mu}_l$ and $\bm{\Sigma}_l$ for $l=1, 2, \dots, L$. Similar to the treatment of $\bm{\alpha}$, we will conditionally maximize the objective function in $(\bm{\mu}_l, \bm{\Sigma}_l)$ for $l=1, 2, \dots, L$ while keeping other variational parameters fixed. Hence, we only describe below how to maximize \cref{eq:Q-in-VI-after-model-params} with respect to $(\bm{\mu}_1, \bm{\Sigma}_1)$.

For the first term in \cref{eq:Q-in-VI-after-model-params}, we apply the commonly used reparameterization technique in VI (see e.g.~\cite{gomes2021insurance}). Let $\bm{w}_{(1)} = (w_{1}^{(1)}, w_{1}^{(2)}, \dots, w_{1}^{(S_1)})^T$ denote the vector form of the first level of random effect $\{w_{1}^{(s)}\}_{s=1, 2, \dots, S_1}$, whose posterior distribution is assumed to be a multivariate standard normal distribution with mean $\bm{\mu}_1$ and diagonal covariance matrix $\bm{\Sigma}_1$. We write $\bm{w}_{(1)} = \bm{\mu}_{1} + \bm{\Sigma}_{1}^{1/2}\bm{v}_{(1)}$ where $\bm{v}_{(1)}$ recovers the standard normal random variables. Let $\bm{v} = \{\bm{v}_{(l)}\}_{l=1, 2, \dots, L}$ denote all such standard normal variables for different levels of random effects.

Assuming $\tilde{\ell}^{c}$ is smooth with finite gradient and hessian, the chain rule for derivatives implies that it is sufficient to calculate only the gradient and hessian of $\tilde{\ell}^{c}$ with respect to $\bm{w}_{(1)}$. For example, the gradient with respect to $\bm{\mu}_1$ can be calculated as
\begin{equation}
\begin{aligned}
    & \frac{\partial}{\partial \bm{\mu}_1}\E_{\bm{w} \sim \bm{q}(\cdot ; \bm{\Theta})} \left[ \tilde{\ell}^{c}(\bm{\alpha}^{(t+1)}, \bm{\beta}^{(t+1)}, \bm{\Psi}^{(t+1)}; \bm{X}, \bm{Y}, \bm{w})\right] \\ 
    &= \E_{\bm{v} \sim \bm{N}(\bm{0}, \bm{I})} \left[ \frac{\partial}{\partial \bm{\mu}_1} \tilde{\ell}^{c}(\bm{\alpha}^{(t+1)}, \bm{\beta}^{(t+1)}, \bm{\Psi}^{(t+1)}; \bm{X}, \bm{Y}, \{ \bm{\mu}_{l} + \bm{\Sigma}_{l}^{1/2}\bm{v}_{(l)} \}_{l=1, 2, \dots, L}) \right] \\
    &= \E_{\bm{w} \sim \bm{q}(\cdot ; \bm{\Theta})} \left[ \frac{\partial}{\partial \bm{w}_{(1)}} \tilde{\ell}^{c}(\bm{\alpha}^{(t+1)}, \bm{\beta}^{(t+1)}, \bm{\Psi}^{(t+1)}; \bm{X}, \bm{Y}, \{\bm{w}_{(l)}\}_{l=1, 2, \dots, L} ) \right]
\end{aligned}
\end{equation}
where the gradient of $\tilde{\ell}^{c}$ with respect to $\bm{w}_{(1)}$ can be obtained by simple differentiation. Recall $\bm{t}_{i1}$ defined in \cref{sec:Mixed-LRMoE-overview} as the zero-one vector that maps policyholder $i$ into one of the levels in the random effect $\bm{w}_{1}$, then
\begin{equation}
\begin{aligned}
    \frac{\partial}{\partial \bm{w}_{(1)}} \tilde{\ell}^{c} &= \sum_{i=1}^{n} \sum_{j=1}^{g} z_{ij}^{(t)} \left( \beta_{j1}^{(t+1)} - \overline{\beta_{1}^{(t+1)}} \right) \times \bm{t}_{i1}
\end{aligned}
\end{equation}
and
\begin{equation}
\begin{aligned}
    \frac{\partial^2}{\partial \bm{w}_{(1)} \partial \bm{w}_{(1)}^T} \tilde{\ell}^{c} &= \sum_{i=1}^{n} \sum_{j=1}^{g} \left\{ z_{ij}^{(t)} \left[  - \overline{(\beta_{1}^{(t+1)})^2} + \left( \overline{\beta_{1}^{(t+1)}} \right)^2 \right] \right\} \times \bm{t}_{i1} \bm{t}_{i1}^T
\end{aligned}
\end{equation}
where
\begin{equation}
    \overline{\beta_{1}^{(t+1)}} = \sum_{j'=1}^g \pi_{j'}(\bm{x}_i, \bm{w}_i; \bm{\alpha}^{(t+1)}, \bm{\beta}^{(t+1)}) \beta_{j'1}^{(t+1)}
\end{equation}
and
\begin{equation}
    \overline{(\beta_{1}^{(t+1)})^2} = \sum_{j'=1}^g \pi_{j'}(\bm{x}_i, \bm{w}_i; \bm{\alpha}^{(t+1)}, \bm{\beta}^{(t+1)}) (\beta_{j'1}^{(t+1)})^2.
\end{equation}
where $\beta_{j1}$ is the first element in $\bm{\beta}_j$ for $j=1, 2, \dots, g$. Similar expressions for the gradient and hessian of $\bm{\Sigma}_{1}$ can also be derived and are omitted. Finally, we update the variational parameters $(\bm{\mu}_1, \bm{\Sigma}_1)$ with a gradient descent formula similar to \cref{eq:IRLS-update-alpha} until the increase in ELBO is negligible, where the gradient and hessian are evaluated through Monte Carlo simulation. The above procedures are then repeated for $l=1, 2, \dots, L$ for all levels of random effects, which yields the updated variational parameters $\bm{\Theta}^{(t+1)} = \{(\bm{\mu}_l^{(t+1)}, \bm{\Sigma}_l^{(t+1)})\}_{l=1, 2, \dots, L}$.

\pagebreak

\bibliographystyle{abbrvnat}
\bibliography{references}  %%% Uncomment this line and comment out the ``thebibliography'' section below to use the external .bib file (using bibtex) .





% \end{thebibliography}

\end{document}

%% file: tables/simulation-1-summary-table.tex
\begin{table}[!ht]
\centering
\caption{Summary of Simulation Study I}
\label{table:simulation-1}

\begin{tabular*}{\textwidth}{c@{\extracolsep{\fill}}c@{\extracolsep{\fill}}c@{\extracolsep{\fill}}}
    \hline \hline
    \multicolumn{3}{l}{\textbf{Coefficients of Fixed Effects (True vs Fitted):}} \\
    & \multicolumn{1}{c}{$\bm{\alpha} = \begin{bmatrix} 1.0 & -1.0 \\ 0 & 0 \end{bmatrix} \quad \hat{\bm{\alpha}} = \begin{bmatrix} 0.8944 & -1.0178 \\ 0 & 0 \end{bmatrix}$} & \\
\end{tabular*}

\begin{tabular*}{\textwidth}{@{\extracolsep{\fill}}c@{\extracolsep{\fill}}cc}
    \multicolumn{3}{l}{\textbf{Expert Functions:}} \\ \hline
    Latent Class & \multicolumn{1}{c}{True} & \multicolumn{1}{c}{Fitted} \\ \hline
    $j=1$ &  \multicolumn{1}{c}{Gamma($k = 2.0, \theta = 1.0$)} & \multicolumn{1}{c}{Gamma($k = 1.9813, \theta = 1.0128$)} \\
    $j=2$ & \multicolumn{1}{c}{Gamma($k = 10.0, \theta = 2.0$)} & \multicolumn{1}{c}{Gamma($k = 10.0819, \theta = 1.9736$)} \\ \hline
\end{tabular*}

\vspace{1em}

\begin{tabular*}{\textwidth}{c@{\extracolsep{\fill}}cccc}
     \multicolumn{5}{l}{\textbf{Recovery of Random Effect} $\{w_{1}^{(s)}\}_{s=1, 2, \dots, 200}$:} \\ \hline
     Posterior CI Level & 90\% & 95\% & 97.5\% & 99\% \\ \hline
     Coverage of Simulated Values & 180 (90\%) & 188 (94\%) & 190 (95\%) & 193 (96.5\%) \\ \hline \hline
\end{tabular*}

\end{table}

%% file: tables/simulation-2-summary-table.tex
\begin{table}[!ht]
\centering
\caption{Summary of Simulation Study II}
\label{table:simulation-2}

\begin{tabular*}{\textwidth}{c@{\extracolsep{\fill}}c@{\extracolsep{\fill}}c@{\extracolsep{\fill}}}
    \hline \hline
    \multicolumn{3}{l}{\textbf{Coefficients of Fixed Effects (True vs Fitted):}} \\
    & \multicolumn{1}{c}{$\bm{\alpha} = \begin{bmatrix} 1.0 & -1.0 \\ 0.75 & 0.25 \\ 0 & 0 \end{bmatrix} \quad \hat{\bm{\alpha}} = \begin{bmatrix} 0.87402 & -0.9327 \\ 0.6221 & 0.3282 \\ 0 & 0 \end{bmatrix}$} & \\
\end{tabular*}

\begin{tabular*}{\textwidth}{c@{\extracolsep{\fill}}c@{\extracolsep{\fill}}c@{\extracolsep{\fill}}}
    \multicolumn{3}{l}{\textbf{Coefficients of Random Effects (True vs Fitted):}} \\
    & \multicolumn{1}{c}{$\bm{\beta} = \begin{bmatrix} 1.0 & 1.0 \\ 0.5 & 1.0 \\ 0 & 0 \end{bmatrix} \quad \quad \hat{\bm{\beta}} = \begin{bmatrix} 1.0 & 1.0 \\ 0.4232 & 1.2586 \\ 0 & 0 \end{bmatrix}$} & \\
\end{tabular*}

\begin{tabular*}{\textwidth}{@{\extracolsep{\fill}}c@{\extracolsep{\fill}}cc}
    \multicolumn{3}{l}{\textbf{Expert Functions:}} \\ \hline
    Latent Class & \multicolumn{1}{c}{True} & \multicolumn{1}{c}{Fitted} \\ \hline
    $j=1$ &  \multicolumn{1}{c}{Gamma($k = 2.0, \theta = 1.0$)} & \multicolumn{1}{c}{Gamma($k = 2.0403, \theta = 0.9704$)} \\
    $j=2$ & \multicolumn{1}{c}{Gamma($k = 20.0, \theta = 1.5$)} & \multicolumn{1}{c}{Gamma($k = 20.3014, \theta = 1.4807$)} \\
    $j=3$ & \multicolumn{1}{c}{Gamma($k = 12.0, \theta = 1.0$)} & \multicolumn{1}{c}{Gamma($k = 11.6926, \theta = 1.0320$)} \\ \hline
\end{tabular*}

\vspace{1em}

\begin{tabular*}{\textwidth}{c@{\extracolsep{\fill}}cccc}
     \multicolumn{5}{l}{\textbf{Recovery of Random Effect} $\{w_{1}^{(s)}\}_{s=1, 2, \dots, 200}$:} \\ \hline
     Posterior CI Level & 90\% & 95\% & 97.5\% & 99\% \\ \hline
     Coverage of Simulated Values & 164 (82\%) & 173 (86.5\%) & 176 (88\%) & 185 (92.5\%) \\ \hline
\end{tabular*}

\vspace{1em}

\begin{tabular*}{\textwidth}{c@{\extracolsep{\fill}}cccc}
     \multicolumn{5}{l}{\textbf{Recovery of Random Effect} $\{w_{2}^{(s)}\}_{s=1, 2, \dots, 2000}$:} \\ \hline
     Posterior CI Level & 90\% & 95\% & 97.5\% & 99\% \\ \hline
     Coverage of Simulated Values & 1536 (76.8\%) & 1672 (83.6\%) & 1765 (88.3\%) & 1858 (92.9\%) \\ \hline \hline
\end{tabular*}

\end{table}

%% file: tables/real-data-overview-table.tex
\begin{table}[!ht]
\centering
\caption{Overview of Real Dataset}
\label{table:real-data-overview}

\begin{tabular*}{\textwidth}{ccl@{\extracolsep{\fill}}}
    \hline \hline
    Covariate & Range & Description \\ \hline
    $x_{i0}$ & 1 & Intercept. Baseline for Female drivers and Rural region. \\
    $x_{i1}$ & $\{0, 1\}$ & Indicator for Male drivers. Mean is 0.47. \\
    $x_{i2}$ & $[16, 99]$ & Driver's age. Mean is 64 and median is 66. \\
    $x_{i3}$ & $[0, 27]$ & Vehicle age. Mean is 6.5 and median is 6. \\
    $x_{i4}$ & $[9.35, 12.82]$ & Natural logarithm of vehicle price. Mean is 10.28 and median is 10.28. \\
    $x_{i5}$ & $[1, 99]$ & Vehicle's collision rating (an indicator of risk). Mean is 26 and median is 27. \\
    $x_{i6}$ & $\{0, 1\}$ & Indicator for policies issued in the Capital. Mean is 0.09. \\
    $x_{i7}$ & $\{0, 1\}$ & Indicator for policies issued in Urban region. Mean is 0.75. \\ \hline \hline
\end{tabular*}

\vspace{1em}

\begin{tabular*}{\textwidth}{c@{\extracolsep{\fill}}c@{\extracolsep{\fill}}c@{\extracolsep{\fill}}c@{\extracolsep{\fill}}c@{\extracolsep{\fill}}c}
    \hline \hline
     Response & $\textrm{Pr}[Y=0]$ & $\textrm{E}[Y|Y>0]$ & $\textrm{SD}[Y|Y>0]$ & $\textrm{Skew}[Y|Y>0]$ & $\textrm{Kurt}[Y|Y>0]$  \\ \hline
     $y_{i}$ (BI) & 0.9790 & 3908 & 2813 & 1.12 & 1.04  \\ \hline \hline
\end{tabular*}

\vspace{1em}

\end{table}

%% file: tables/real-data-model-benchmark.tex
\begin{table}[!ht]
\centering
\caption{Benchmark Models for Real Data Analysis}
\label{table:real-data-model-benchmark}

\begin{tabular*}{\textwidth}{l@{\extracolsep{\fill}}l@{\extracolsep{\fill}}l@{\extracolsep{\fill}}c@{\extracolsep{\fill}}c@{\extracolsep{\fill}} c}
    \hline \hline
    Benchmark &  Claim Probability & log(Claim Severity) & loglik (Training) & loglik (Testing) & \#(Parameters)\\ \hline
    GLM-GLM & Logistic GLM & Gaussian GLM & $-22842.70$ & $-3260.98$ & 16 \\
    GLM-GLMM & Logistic GLM & Gaussian GLMM & $-22781.38$ & $-3258.38$ & 17 \\
    GLMM-GLM & Logistic GLMM & Gaussian GLM & $-22578.33$ & $-3260.22$ & 17 \\
    GLMM-GLMM & Logistic GLMM & Gaussian GLMM & $-22517.01$ & $-3257.62$ & 18 \\\hline \hline
\end{tabular*}

\end{table}

%% file: tables/real-data-model-LRMoEs.tex
\begin{table}[!ht]
\centering
\caption{(Mixed) LRMoE Models for Real Data Analysis}
\label{table:real-data-model-LRMoEs}

\begin{tabular*}{\textwidth}{c@{\extracolsep{\fill}}c@{\extracolsep{\fill}}c@{\extracolsep{\fill}}c@{\extracolsep{\fill}}c@{\extracolsep{\fill}}c@{\extracolsep{\fill}}c}
    \hline \hline
    & & \multicolumn{2}{c}{Training} & \multicolumn{2}{c}{Testing} & \\ \cline{3-4} \cline{5-6}
    Model & $g$ & loglik & ELBO & loglik & ELBO & \#(Parameters) \\ \hline
    LRMoE (4) & 4 & $-22688.30$ &  & $-3249.91$ & & 32 \\ % 9-4
    Mixed LRMoE (4) & 4 & $-22098.66$ & $-23454.49$ & $-3248.07$ & $-3922.89$ & 34 \\
    Mixed LRMoE (5) & 5 & $-22081.38$ & $-23464.97$ & $-3246.98$ & $-4002.94$ & 45 \\ \hline \hline 
\end{tabular*}

\end{table}

%% file: tables/real-data-prob-changes.tex
\begin{table}[!ht]
% \centering
\caption{Comparison of latent classes and the predicted probabilities by claim history. The first table summarizes the mean and standard deviation of the response by latent class, and we have manually categorized them into three risk levels. The second table compares the predicted latent class probabilities for different groups of policyholders by their claim history (No: no claim during 2014--2018. Yes: at least one claim during 2014--2018), calculated from different models.}
\label{table:real-data-prob-changes}

\vspace{0.5em}

\begin{tabular*}{\textwidth}{c@{\extracolsep{\fill}}ccc@{\extracolsep{\fill}}ccc@{\extracolsep{\fill}}ccc}
    \hline \hline
    &  \multicolumn{3}{c}{LRMoE (4)}      &  \multicolumn{3}{c}{Mixed LRMoE (4)}     &  \multicolumn{3}{c}{Mixed LRMoE (5)}     \\ \cline{2-4} \cline{5-7} \cline{8-10}
Risk Level & Class    & Mean  &  SD & Class    & Mean  &  SD & Class    & Mean  &  SD  \\ \hline
Low & 1 & 0.41 & 11 & 1 & 0.08 & 5 & 1 & 0 & 0 \\
& & & & 2 & 0.58 & 14 & 2 & 0.83 & 16 \\ \hline
Medium & 2 & 84 & 879 & 3 & 54 & 719 & 3 & 25 & 469 \\
& 3 & 130 & 756 & & & & 4 & 27 & 535 \\ \hline
High & 4 & 269 & 1194 & 4 & 1365 & 2431 & 5 & 1476 & 2539 \\ \hline \hline
\end{tabular*}

\vspace{1em}

\begin{tabular*}{\textwidth}{c@{\extracolsep{\fill}}cc@{\extracolsep{\fill}}cc@{\extracolsep{\fill}}cc}
    \hline \hline
    &  \multicolumn{2}{c}{LRMoE (4)}      &  \multicolumn{2}{c}{Mixed LRMoE (4)}     &  \multicolumn{2}{c}{Mixed LRMoE (5)}     \\ \cline{2-3} \cline{4-5} \cline{6-7}
Risk Level & No    & Yes  &  No        & Yes  &  No        & Yes  \\ \hline
Low & 0.37 & 0.33 & 0.68 & 0.63 & 0.44 & 0.40 \\
Medium & 0.55 & 0.58 & 0.28 & 0.27 & 0.51 & 0.51 \\
High & 0.08 & 0.09 & 0.05 & 0.10 & 0.05 & 0.09 \\ \hline \hline
\end{tabular*}

\vspace{1em}

\end{table}

%% file: tables/real-data-premium-table.tex
\begin{table}[!ht]
% \centering
\caption{Average of predicted posterior premium based on policyholders' claim history. The cutoff points for positive claim sizes are the 33\% and 67\% percentiles of its distribution. Percentages in brackets indicate the additional premium loadings compared with policyholders without any claim history, i.e.~Claim Indicator = No.}
\label{table:real-data-comparison-premium-table}

\vspace{0.5em}

\begin{tabular*}{\textwidth}{l@{\extracolsep{\fill}}cc@{\extracolsep{\fill}}cccc}
    \hline \hline
    & \multicolumn{2}{c}{Claim Indicator} & & \multicolumn{3}{c}{Claim Size} \\ \cline{2-3} \cline{5-7}
    Model           & No    & Yes          & & Small       & Medium          & Large     \\ \hline
GLM-GLM         & 78 & 87 (13\%)  & & 86 (11\%)  & 86 (11\%)  & 90 (16\%)  \\
GLM-GLMM        & 87 & 97 (13\%)  & & 96 (11\%)  & 96 (11\%)  & 100 (16\%) \\
GLMM-GLM        & 78 & 103 (32\%) & & 101 (31\%) & 101 (31\%) & 105 (36\%) \\
GLMM-GLMM       & 87 & 114 (32\%) & & 113 (31\%) & 113 (31\%) & 117 (36\%) \\
LRMoE (4)       & 82 & 88 (7\%)   & & 88 (7\%)   & 87 (6\%)   & 89 (9\%)   \\
Mixed LRMoE (4) & 82 & 151 (85\%) & & 147 (81\%) & 153 (88\%) & 152 (87\%) \\
Mixed LRMoE (5) & 86 & 158 (84\%) & & 155 (81\%) & 158 (85\%) & 160 (87\%) \\ \hline \hline
\end{tabular*}

\vspace{1em}

\end{table}

%% file: tables/real-data-Gini-comparison.tex
\begin{table}[!ht]
% \centering
\caption{Summary of Gini Index. Numbers in brackets indicate the estimated standard error.}
\label{table:real-data-comparison-Gini}

\vspace{0.5em}

\textbf{Training Set}:

\vspace{0.5em}

\begin{tabular*}{\textwidth}{c@{\extracolsep{\fill}}c@{\extracolsep{\fill}}c@{\extracolsep{\fill}}c@{\extracolsep{\fill}}c}
    \hline \hline
    Model & GLM-GLM & GLM-GLMM & GLMM-GLM & GLMM-GLMM \\ \hline
    Gini Index & 0.2191 (0.0165) & 0.2191 (0.0165) & 0.4292 (0.0154) & 0.4292 (0.0154) \\ \hline
    Model & LRMoE (4) & Mixed LRMoE (4) & Mixed LRMoE (5) &  \\ \hline
    Gini Index & 0.2163 (0.0163) & 0.7002 (0.0130) & 0.7266 (0.0123) &  \\ \hline \hline
\end{tabular*}

\vspace{1em}

\textbf{Testing Set}:

\vspace{0.5em}

\begin{tabular*}{\textwidth}{c@{\extracolsep{\fill}}c@{\extracolsep{\fill}}c@{\extracolsep{\fill}}c@{\extracolsep{\fill}}c}
    \hline \hline
    Model & GLM-GLM & GLM-GLMM & GLMM-GLM & GLMM-GLMM \\ \hline
    Gini Index & 0.2999 (0.0415) & 0.2999 (0.0415) & 0.3057 (0.0421) & 0.3057 (0.0421) \\ \hline
    Model & LRMoE (4) & Mixed LRMoE (4) & Mixed LRMoE (5) &  \\ \hline
    Gini Index & 0.3030 (0.0403) & 0.2892 (0.0411) & 0.2880 (0.0413) &  \\ \hline \hline
\end{tabular*}

\vspace{1em}

\end{table}